%% file: main.tex
\documentclass[twoside,journal]{IEEEtran}

\usepackage{cite}
\usepackage{amsmath,amssymb,amsfonts}
\usepackage{algorithm}
\usepackage{algpseudocode}
\usepackage{array}
\usepackage{graphicx}
\usepackage{textcomp}
\usepackage{url}
\usepackage{booktabs}
\usepackage{threeparttable}
\usepackage{multirow}
\usepackage[table]{xcolor}
\usepackage{arydshln}
\usepackage{bm}
\usepackage{amsthm}
\newtheorem{definition}{Definition}
\newtheorem{proposition}{Proposition}
\newtheorem{theorem}{Theorem}
\newtheorem{lemma}{Lemma}

\usepackage[caption=false,font=normalsize,labelfont=sf,textfont=sf]{subfig}

\newcommand{\SpeedupLo}{1.51}
\newcommand{\SpeedupHi}{1.92}
\newcommand{\SpeedupGbdt}{2}
\usepackage{hyperref}
\usepackage{orcidlink}
\usepackage{mathtools}

\usepackage{pdfpages}
\begin{document}

\title{Signed Dyadic Path Encoding for Efficient Non-Interactive Private Decision Tree Evaluation over TFHE}

\author{Xintong Song\orcidlink{0009-0007-9578-3878},
    Yiqiang Zhao\orcidlink{0000-0002-4564-952X},
    Qizhi Zhang\orcidlink{0000-0002-9111-119X},
    Yao Li\orcidlink{0000-0002-7014-8145},
    Jiaji He\orcidlink{0000-0003-1443-9279},
    and Fei Zhao\orcidlink{0009-0000-1176-7301}
    \thanks{This work has been submitted to the IEEE for possible publication. Copyright may be transferred without notice, after which this version may no longer be accessible.}
    \thanks{This work was supported by Science Challenge Project, No. TZ2025002. \textit{(Corresponding author: Fei Zhao.)}}
    \thanks{Xintong Song, Yiqiang Zhao, Qizhi Zhang, Yao Li, and Jiaji He are with the School of Microelectronics, Tianjin University, Tianjin 300372, China (email: \url{xintong_song@tju.edu.cn}, \url{yq_zhao@tju.edu.cn}, \url{qizhi_zhang@tju.edu.cn}, \url{liyao@tju.edu.cn}, \url{dochejj@tju.edu.cn}).}
    \thanks{Fei Zhao is with the State Key Laboratory of Multimedia Information Processing, School of Computer Science, Peking University, No.5 Yiheyuan Road, Haidian District, Beijing 100871, China (email: \url{feizhao@pku.edu.cn}).}
}

\maketitle

\bstctlcite{IEEEexample:BSTcontrol}

\begin{abstract}
Private decision-tree evaluation lets a server classify a client's encrypted query without learning the input.
In the non-interactive, path-based setting over TFHE, the server returns a single encrypted prediction by aggregating per-path comparison outcomes, and the standard per-leaf formulation pays one programmable bootstrap (PBS) per leaf, which dominates the cost.
We propose a signed dyadic path-sum encoding that nearly halves this cost: each internal node receives a signed dyadic label so that every root-to-leaf path accumulates a distinct integer sum, which lets the two children of a leaf-parent be packed into a single test polynomial and resolved by one PBS, reducing the PBS count from the number of leaves to the number of leaf-parents.
We formalize the distinct-sum property and prove that the dyadic labeling is the range-minimal encoding satisfying it, and its half-symmetric signed form is magnitude-optimal, tightening the comparison-stage noise bound at fixed parameters.
We derive the two conditions under which the encrypted evaluation returns the exact prediction and instantiate parameters with per-inference failure below \(2^{-64}\) at 128-bit security.
We show that the ring degree grows linearly with the consumed plaintext range, which confines the profitable regime to depths \(\leq 4\), where our method stays faster even when each scheme runs at its own cost-optimal parameters.
We further extend the encoding to gradient-boosted ensembles.
On standard UCI benchmarks at depths \(1\)--\(4\), our method reduces single-tree latency by \(\SpeedupLo\times\)--\(\SpeedupHi\times\) and ensemble latency by \(\SpeedupGbdt\times\), without accuracy loss, while preserving the compact, non-interactive response.
\end{abstract}


\section{Introduction}\label{sec:introduction}
\input{introduction}

\section{Preliminaries}\label{sec:preliminaries}
\input{preliminaries}

\section{Proposed Method}\label{sec:method}
\input{method}

\section{Experimental Evaluation}\label{sec:experiment}
\input{experiment}

\section{Discussion}\label{sec:discuss}
\input{discussion}

\section{Conclusion}\label{sec:conclusion}
\noindent Deploying privacy-preserving inference on lightweight platforms requires efficient non-interactive tree evaluation.
This work nearly halves that cost through a change confined entirely to the encoding.
By assigning each tree level a signed dyadic label and accumulating a distinct path sum for every RTL path, the two children of each leaf-parent are packed into a single lookup table, allowing one PBS to both identify the active child and return its label.
The PBS count thereby falls from the number of leaves \(|\mathcal{L}|\) to the number of leaf-parents \(|\mathcal{P}|\).
We proved that the half-symmetric signed system \(\mathcal{S}^*\) minimizes the worst-case path-sum magnitude.
We also showed that it tightens the comparison-stage noise bound relative to the unsigned dyadic labeling, and derived parameters with an estimated per-inference failure probability below \(2^{-64}\) under the standard noise model, at \(128\)-bit security.
Our method is an interface-compatible replacement for the state-of-the-art since it computes the exact plaintext result and preserves the single-ciphertext, non-interactive response.

On four UCI benchmarks at depths \(1\)--\(4\), with each scheme run at its own cost-optimal parameters, our method reduces single-tree inference latency by \(\SpeedupLo\times\)--\(\SpeedupHi\times\), and it accelerates a depth \(3\), \(80\) trees per class gradient boosting forest by \(\SpeedupGbdt\times\).
Two directions promise to broaden the construction's reach: tightening the modulus-switching floor via drift-aware failure analysis~\cite{BJSW25} or reduced-noise PBS, and extending the encoding to multi-ary trees.
Together, these advance a broader goal: bringing efficient, non-interactive PDTE to the lightweight, resource-constrained platforms where trustworthy inference matters most.

\bibliographystyle{IEEEtran}
\bibliography{references}

\clearpage
\includepdf[pages=-]{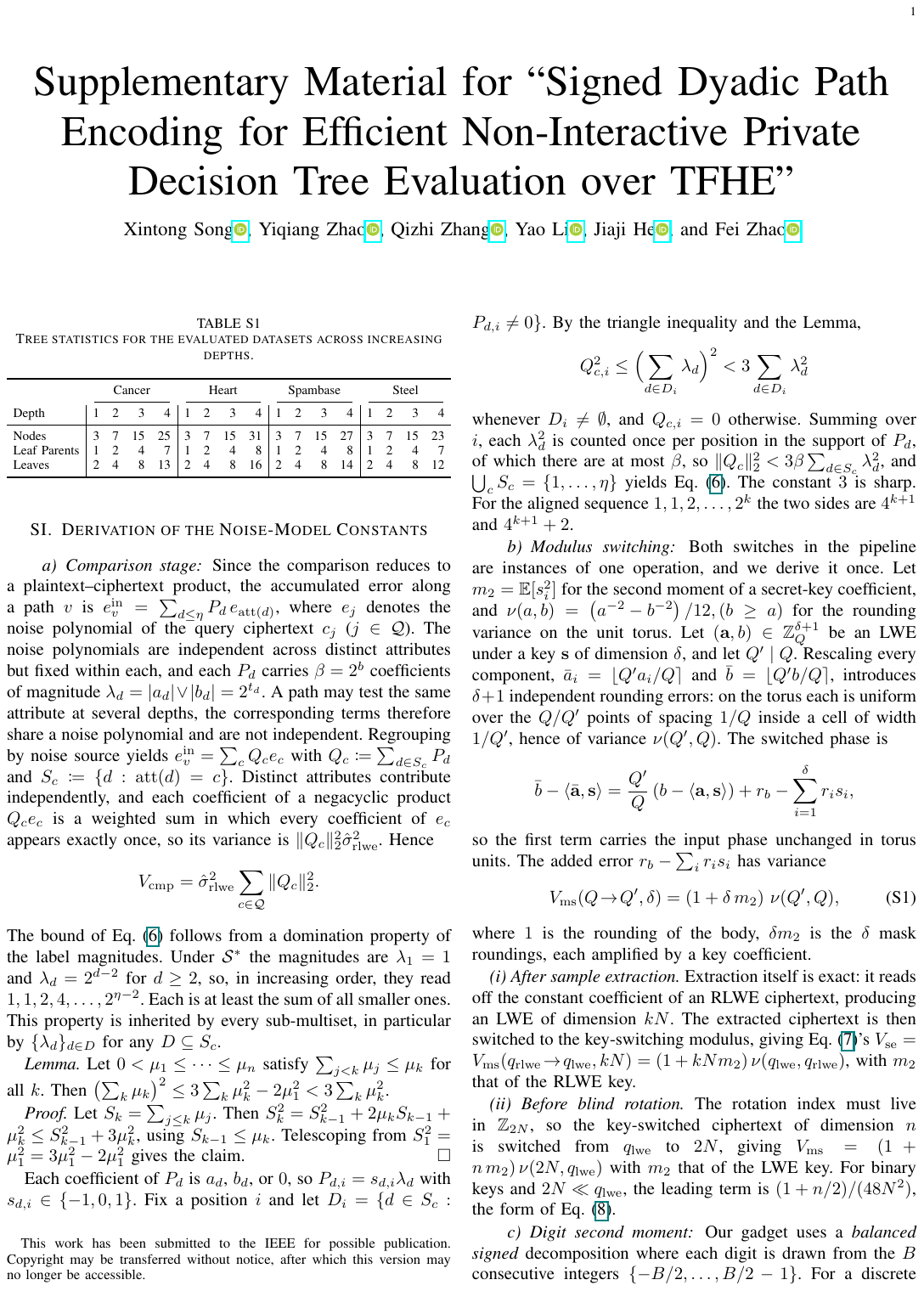}
\end{document}

%% file: introduction.tex
\begin{figure}[t]
    \centering
    \includegraphics[width=\columnwidth]{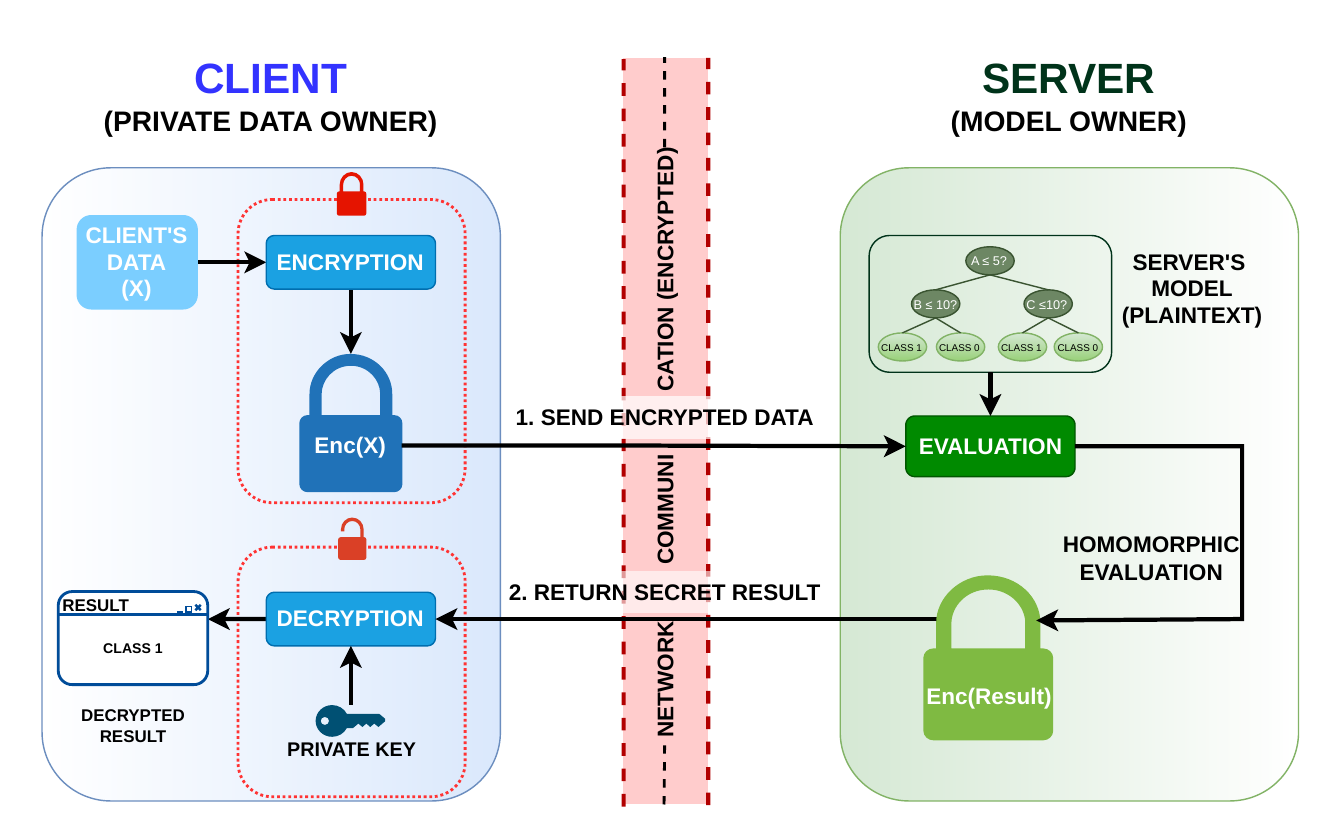}
    \caption{Overview of non-interactive PDTE under homomorphic encryption (HE). The client sends its encrypted private feature vector to the server, which holds the decision tree model in plaintext. The server homomorphically evaluates the tree without decryption and returns the encrypted result, which only the client can decrypt with its private key to obtain the final classification.}
    \label{fig:pdte}
\end{figure}

\IEEEPARstart{M}{achine}-Learning-as-a-Service lets clients obtain predictions from models hosted on remote servers, yet it compels them to disclose sensitive inputs, such as medical records, financial histories, behavioral traces, to the service provider, placing privacy-preserving inference at the core of trustworthy machine-learning deployment.
Decision trees and their ensembles remain the state-of-the-art for classification on tabular data~\cite{GOV22}, making private inference over tree-based models a practical challenge.
Private decision tree evaluation (PDTE) addresses it by allowing a server to classify a client's query under a decision tree while guaranteeing that the server learns nothing about the input, as depicted in Figure~\ref{fig:pdte}.
The feasibility of this guarantee is by now well established.
The open frontier, and the subject of this work, is realizing it efficiently enough for servers with intermittently connected clients under a minimal per-query uplink.

PDTE protocols can be broadly categorized into two principal paradigms. \emph{Interactive} protocols, which are built on secure multi-party computation or garbled circuits, achieve strong computational efficiency.
However, they require the client and server to exchange messages over multiple rounds, with communication that typically scales with tree depth, necessitating both parties remain online throughout the computation~\cite{BPTG15, WFNL16, TMZC17, KNLA19, MTZC21, ADD22, YLQ+24, KMYO+24, FWO25, XZZ+26}.
In contrast, \emph{non-interactive} protocols leverage cryptographic primitives such as HE to enable a different trade-off: the client encrypts its query, transmits it in a single message, and may then go offline while the server evaluates the entire decision tree under encryption, ultimately returning a single ciphertext from which the client decrypts the prediction~\cite{LZS18, XWCF19, TBK20, SNR+21, ADGK22, CDPP22, PTVA22, MNLK23, FSB+23, CKNP24, LLG+24, DABY26}.
This non-interactive model is particularly attractive when the client is bandwidth-limited, intermittently connected, or computationally constrained, and it constitutes the setting investigated in this work.

Path-based PDTE offers an optimal combination of performance and communication, with MultiplyPath and SumPath as the two principal approaches.
The query budget, however, is not a free parameter.
The MultiplyPath family~\cite{BPTG15,WFNL16,CDPP22} can evaluate the tree without any PBS, but its traversal consumes RGSW-format operands, which must be obtained in one of three ways: shipped by the client as \(\ell\) gadget-scaled ciphertexts per attribute, inflating the uplink by a factor of \(\ell\); reconstructed on the server via circuit bootstrapping, at a cost of approximately \(\ell\) PBS per internal node; or recovered through transciphering, which effectively restores a per-node PBS traversal.
Each alternative trades the compact query for server cost, or vice versa.
The SumPath family~\cite{TMZC17,KNLA19,MNLK23} preserves a minimal client query of one RLWE ciphertext per attribute.
In a TFHE realization, however, its leaf selection requires a zero-test per leaf (TE0), implemented as a PBS that tests, for each leaf, whether its path sum equals zero.
Because a PBS is by far the most expensive TFHE primitive, leaf selection incurs \(|\mathcal{L}|\) PBS operations for a tree with \(|\mathcal{L}|\) leaves and dominates the inference time.

The limiting factor is therefore not feasibility but the cost of leaf selection under a minimal query budget.
This raises the central question we address: under a fixed uplink of one RLWE ciphertext per attribute, can it be accomplished with strictly fewer PBSs, without degrading accuracy or the compact single-ciphertext response profile?
We answer in the affirmative by proposing an encoding scheme that resolves the two leaves sharing a common parent within a single PBS, reducing the PBS count by a factor up to \(2\times\) for general tree topologies, while preserving the minimal query and response profile.

\subsection{Related Work}\label{subsec:related-work}

\subsubsection{Private Comparison}
A substantial body of work concentrates on the comparison stage, developing high-precision operators such as XCMP~\cite{LZS18}, which achieves constant multiplication depth, and its successor XXCMP ~\cite{MNLK23}, which extends the approach to support higher-precision data.
Subsequent designs, including range-cover comparison (RCC) ~\cite{MNLK23,CKNP24} and the RDCMP comparator~\cite{LLG+24}, improve amortized performance by enabling batched comparisons across multiple inputs.

\subsubsection{Tree Traversal}
Modern approaches for private tree traversal can be largely categorized into four types: \emph{SinglePath, MatrixTransform, MultiplyPath, and SumPath}.

\noindent\textbf{SinglePath}
The SinglePath method~\cite{ADGK22,XTWA23,YLQ+24} traverses the tree in a level-by-level fashion.
For instance, ~\cite{ADGK22} introduced a Blind Node Selection technique that selects the next node to traverse based on the encrypted comparison result, thereby evaluating a single root-to-leaf (RTL) path.
The primary advantage of these protocols is a reduction in the number of private comparisons, which equals only the depth of the tree.
The drawback, however, is that the server must perform a Blind Array Access operation to determine which input to compare at each level, since the outcome of the preceding comparison remains encrypted and unknown to the server.
This operation is expensive under HE primitives, making these protocols less attractive in the single-query setting.
\cite{PCL26} revisits this paradigm over CKKS.
It combines one-branch-only comparison with baby-step giant-step branch selection to reduce the server cost from \(O(2^{\eta})\) to \(O\left(p\cdot\sqrt{2^{\eta}}\right)\).
Its gains are amortized over a full batch of SIMD slots and appear at larger depths where \(\eta\ge7\) in its evaluation.
This leaves the single-query shallow-tree regime unaddressed.

\noindent\textbf{MatrixTransform}
The MatrixTransform method~\cite{FSB+23,FWO25} traverses the tree using tensor-based operations.
These protocols encode the decision tree as a feature selection matrix, where each row corresponds to a distinct node.
During traversal, the tree matrices are multiplied by a quantized encrypted data vector and the results are summed along each RTL path.
PBS comparisons are then performed for each leaf to produce a one-hot vector.
This vector is subsequently multiplied by a leaf-value matrix to yield the final encrypted, one-hot-encoded result vector.
The principal advantage of these protocols lies in their ability to fully parallelize execution across many nodes.
However, the rigid matrix representation makes them less suitable for irregular trees, where it can introduce substantial computational overhead~\cite{DABY25}.

\noindent\textbf{MultiplyPath}
The MultiplyPath method~\cite{BPTG15,WFNL16,CDPP22}, also referred to as PathConjugation, traverses the tree by performing a chained ciphertext multiplication along each RTL path.
In these protocols, the comparison operation assigns a value of one to the selected child node and zero to the other.
Consequently, after the chained multiplication, only the leaf on the active path retains a value of one, while all others become zero.
Each leaf is then multiplied by its corresponding leaf value, and the final result is obtained by summing over all leaves.
The advantages of this approach are its fully parallelizable nature and the simplicity of the leaf aggregation logic, which reduces to an \(O(|\mathcal{L}|)\) homomorphic addition.
However, as discussed previously, it incurs a fundamental trade-off between server computation and client query size.

\noindent\textbf{SumPath}
The SumPath construction~\cite{TMZC17,KNLA19,MNLK23,CKNP24,SCL+24,DABY26} traverses the tree by summing the node values along each RTL path.
In these protocols, a comparison is performed at each internal node, assigning a value \(t \in \{0,1\}\) to one child and \(1-t\) to the other.
After summation along each path, only the active path yields a sum of zero, while all other paths produce non-zero values.
Subsequently, a zero-test procedure is applied on the server to isolate the correct leaf.
The resulting indicator is then multiplied by the leaf values and aggregated, returning a single encrypted result to the client.
Like MatrixTransform and MultiplyPath, SumPath is fully parallelizable.

\subsubsection{Client- vs.\ server-side leaf selection}
How the path results are resolved into a prediction divides existing schemes into two categories. \emph{Client-side selection} returns the encrypted path results to the client, which decrypts them and identifies the active leaf locally~\cite{TMZC17,KNLA19,MNLK23}.
This approach keeps the server-side workload low but precludes server-side aggregation.
The client must reconstruct the prediction on its own, and ensemble evaluation cannot be completed without revealing per-tree information.
\emph{Server-side selection} resolves the active leaf entirely on the server and returns a single ciphertext.
The aforementioned TE0 procedure~\cite{CKNP24} is the representative technique.
It maps the zero-sum path to one and all other paths to zero, then combines the leaf values into a single ciphertext, enabling full non-interactivity and seamless ensemble support.
However, this routine is inherently expensive.
On TFHE, it requires one PBS per leaf, while on BFV/CKKS it performs \(\eta - 1\) ciphertext multiplications per leaf and one relinearization per multiplication.
Therefore, this stage dominates the inference cost and grows exponentially with tree depth.

\subsubsection{Batched amortization and its trade-offs}
To mitigate this cost, recent work exploits the SIMD packing capabilities of BFV/CKKS to batch many queries simultaneously, amortizing the leaf selection overhead and improving throughput~\cite{KMYO+24,CKNP24,LLG+24,PCL26}.
This approach, however, either leaves the single-query latency unaffected or reintroduces an interaction round to realize the selection, weakening the non-interactive guarantee.

TFHE, by contrast, natively supports single-query, fully non-interactive evaluation suited to offline and lightweight clients~\cite{CDPP22}.
Because the ring degree is governed by the PBS rather than by the multiplicative depth of the circuit, it keeps the working parameters small and the response a single compact ciphertext.
But its per-leaf PBS remains the unaddressed bottleneck~\cite{KMYO+24}.

\subsubsection{Gap and our direction}\label{subsubsec:gap}
Table~\ref{tab:complexity} places our construction among the schemes that share our regime, namely a single query message, server-side resolution of the active leaf, and a single-ciphertext response under TFHE. Restricting the comparison to TFHE realizations makes the blind rotation a common and platform-independent unit, whereas measured latencies across works are determined by implementation and parameter choices that differ in every case.
The relevant distinction is where the cost is paid, not how much.
SortingHat sits at the MultiplyPath extreme of this trade.
Sending \(\ell\) gadget-scaled RLWE ciphertexts per attribute, which the server converts to RGSW by key switching, yields a PBS-free traversal at an \(\ell\)-fold uplink, while the server-side reconstruction routes recover the compact query only by restoring a per-node PBS cost.
The trade is symmetric at the other end where an LWE query is more compact, but places the node comparison inside a PBS, so the feature precision enters the cost of every blind rotation.

Within the minimal-query class, leaf selection costs one blind rotation per leaf, and schemes differ only in what they add at the node stage, with~\cite{FSB+23} and~\cite{PTVA22} spending a further \(|\mathcal{I}|\) PBSs on the comparison and on the complement respectively.
TE0 is therefore the least expensive member of the class and the appropriate one to compare.
Existing schemes reduce \(|\mathcal{L}|\) only by removing zero-valued leaves~\cite{CKNP24}.
None packs the active-leaf test for sibling leaves.
We close this gap by replacing the per-leaf zero-tests with a single packed PBS that resolves both children of a leaf-parent at once.
This lowers the PBS count from \(|\mathcal{L}|\) to \(|\mathcal{P}|\), while the signed labels supply the branch indicator directly and remove the complement from the node stage.
The reduction introduces no evaluation error and leaves the query, the response, and the privacy guarantee unchanged.

\begin{table*}[!t]
    \centering
    \begin{threeparttable}
        \caption{Non-interactive, single-response TFHE schemes: per-inference cost. \(\mathcal{I}\), \(\mathcal{P}\), \(\mathcal{L}\) denote the sets of internal nodes, leaf-parents, and leaves, \(t\) is the operand bitwidth, \(\ell\) the gadget level, and \(N\) the ring degree.}
        \label{tab:complexity}
        \begin{tabular}{llcccc}
            \toprule
            Scheme & Family & Query / attribute & Blind rotations & Other server cost & Response \\
            \midrule
            Frery \emph{et al.}~\cite{FSB+23} & MatrixTransform & 1 LWE & \(|\mathcal{I}| + |\mathcal{L}|\)  & matrix products & 1 LWE \\
            SortingHat~\cite{CDPP22}          & MultiplyPath    & \(\ell\) RLWE\tnote{3} & 0\tnote{3}  & \(|\mathcal{I}|(\log N + 1)\) ext.\ products & 1 RLWE \\
            Paul \emph{et al.}~\cite{PTVA22}  & SumPath         & \(\lceil t/4 \rceil\) LWE    & \(|\mathcal{I}| + |\mathcal{L}|\) & prefix sums & 1 LWE \\
            TE0 (TFHE)~\cite{CKNP24}\tnote{4} & SumPath   & 1 RLWE & \(|\mathcal{L}|\) & \(|\mathcal{I}|\) (CP + CPE + SUM) & 1 LWE \\
            \textbf{Ours}                     & SumPath   & 1 RLWE & \(\boldsymbol{|\mathcal{P}|}\) & \(|\mathcal{I}|\) (CP + SUM) & 1 LWE \\
            \bottomrule
        \end{tabular}
        \begin{tablenotes}
            \item[1] Measured latency is not comparable across implementations and platforms and is therefore not reported. The controlled comparison against TE0 is given in Section~\ref{sec:experiment}.
            \item[2] CP: node comparison by polynomial multiplication. CPE: node complement. SUM: RLWE addition.
            \item[3] The PBS-free traversal holds only when the client sends \(\ell\) gadget-scaled RLWE ciphertexts per attribute, which the server converts to RGSW by key switching, at an \(\ell\)-fold uplink. The remaining routes reintroduce blind rotations: circuit bootstrapping costs \(\approx \ell\) per internal node, and transciphering restores a per-node PBS traversal.
            \item[4] Originally a leveled-HE construction. Counts are for our TFHE realization (Section~\ref{subsec:baseline}).
        \end{tablenotes}
    \end{threeparttable}
\end{table*}

\subsection{Our Contributions}\label{subsec:our-contributions}

We make the following contributions:

\begin{enumerate}
    \item \textbf{A leaf-selection primitive that nearly halves the PBS count.} We propose a signed dyadic path-sum encoding that packs the two children of each leaf-parent into a single lookup table (LUT), reducing the leaf-selection cost from one PBS per leaf \(|\mathcal{L}|\) to one PBS per leaf-parent \(|\mathcal{P}|\).
    The same encoding streamlines path evaluation: each node carries two precomputed dyadic labels selected directly by the comparison bit, which removes the per-node complement of classical SumPath and accumulates only to leaf-parents rather than to all leaves.
    The primitive is an interface-compatible replacement for the prior TE0 aggregation, computing the exact plaintext result while preserving the compact single-ciphertext response per tree.

    \item \textbf{An optimality theory for the encoding.} We formalize the distinct-sum property as a condition on integer label systems and prove that a dyadic labeling essentially uniquely minimizes the path-sum range.
    We further prove that the half-symmetric signed dyadic system minimizes the worst-case path-sum magnitude.
    The smaller label magnitudes also tighten the comparison-stage noise bound, providing additional decryption margin that becomes material in comparison-noise-dominated systems.

    \item \textbf{A construction with an explicit failure analysis, evaluated against the state of the art.} We prove correctness of the encrypted evaluation, identifying the PBS input and aggregated output conditions that govern TFHE-based SumPath PDTE generally.
    We instantiate parameters whose estimated per-inference failure probability, under the standard independence heuristic, is below \(2^{-64}\) at \(128\)-bit security.
    We further characterize the depth trade-off: because the required ring degree grows linearly with the consumed plaintext range, the depth ceiling is an economic boundary, and the profitable regime covers the shallow depths (\(\leq 4\)) typical of practical tabular models.
    Implementing both our primitive and the most efficient TFHE realization of the prior TE0 aggregation in a common pipeline, we compare them with each scheme run at its own cost-optimal parameters to achieve a fair, system-level comparison.
    On four standard UCI datasets at depths \(1\)--\(4\), our method reduces single-tree inference latency by \(\SpeedupLo\times\)--\(\SpeedupHi\times\) over TE0, without accuracy loss, narrowing to \(1.5\times\)--\(1.8\times\) only at the deepest profitable depth, where the baseline's smaller range earns it a lighter configuration.

    \item \textbf{Extension to gradient-boosted ensembles.} We extend the encoding to gradient boosting decision trees (GBDT), deriving the rescaling factor and torus parameters that keep the aggregated evaluation correct under the target failure probability.
    On a multi-class dataset, a depth \(3\) ensemble of \(80\) trees per class improves model accuracy by \(18\) percentage points over the single-tree baseline, and our method reduces latency by \(\SpeedupGbdt\times\) against the same baseline with no loss of accuracy.
    These results confirm that the reduction in PBS count translates directly into end-to-end speedup in the non-interactive, single-query regime for practical gradient-boosted evaluation.
\end{enumerate}

The remainder of the paper is organized as follows.
Section~\ref{sec:preliminaries} reviews TFHE and the necessary background.
Section~\ref{sec:method} presents our encoding and its optimality.
Section~\ref{sec:experiment} reports the experimental evaluation.
Section~\ref{sec:discuss} discusses extensions and open questions, and Section~\ref{sec:conclusion} concludes.

%% file: preliminaries.tex
\subsection{Notation}\label{subsec:notation}
Let \(\mathbb{T} = \mathbb{R}/\mathbb{Z}\) denote the real torus, \(\mathbb{B} = \{0,1\}\), and let \(N\) be a power of two.
We write \(\mathbb{Z}_N[X] = \mathbb{Z}[X]/(X^N+1)\) and \(\mathbb{T}_N[X] = \mathbb{T}[X]/(X^N+1)\) for the corresponding negacyclic polynomial rings.
Vectors are denoted in bold as \(\mathbf{a}\), and the inner product is written as \(\langle \mathbf{a}, \mathbf{s}\rangle\).
A discrete message \(m \in \mathbb{Z}_p\), where \(p\) is the \emph{plaintext modulus}, is embedded into the torus as \(m/p \in \mathbb{T}\).
All schemes described below are instantiated over a discretized torus \(\mathbb{Z}_q\). \(\eta\) denotes the tree depth, and a node at level \(d \in \{1,\dots,\eta\}\) is one whose path from the root consists of \(d-1\) edges.
Table~\ref{tab:notation} summarizes the notation used throughout this paper.

\begin{table}[t]
    \centering
    \caption{Notation used throughout this paper.}
    \label{tab:notation}
    \begin{tabular}{ll}
        \toprule
        Symbol & Meaning \\
        \midrule
        \(\mathcal{T}\) & decision tree \\
        \(\mathcal{I}\) & set of internal nodes \\
        \(\mathcal{P}\) & set of leaf-parents \\
        \(\mathcal{L}\) & set of leaf nodes\\
        \(\eta\) & tree depth \\
        \(d\) & node level, \(d\in\{1,\dots,\eta\}\) \\
        \(\varepsilon_d(\alpha, \beta)\) & branch indicator at level \(d\) (\(\alpha\) left, \(\beta\) right) \\
        \((a_d,b_d)\) & level-\(d\) label (contribution \(a_d\)/\(b_d\) for left/right) \\
        \(\delta_d\) & increment \(b_d-a_d\) \\
        \(\sigma_v\) & cumulative path sum to node \(v\) \\
        \(\tau_v\) & threshold value of a node \(v\) \\
        \(\Sigma(\boldsymbol{\varepsilon})\) & sum map of a configuration \(\boldsymbol{\varepsilon}\) \\
        \(M(\mathcal{S})\) & worst-case absolute path sum \(\max_{\boldsymbol{\varepsilon}}|\Sigma(\boldsymbol{\varepsilon})|\) \\
        \(f_p\) & packed test polynomial for leaf-parent \(p\) \\
        \(N\) & RLWE dimension (LUT polynomial degree) \\
        \(n\) & LWE dimension \\
        \(p\) & plaintext modulus \\
        \bottomrule
    \end{tabular}
\end{table}

\subsection{TFHE Ciphertexts}\label{subsec:tfhe}
TFHE~\cite{CGGI20} defines three ciphertext types over the torus, each serving a distinct role in homomorphic computation.
A \emph{TLWE} ciphertext encrypts a scalar message \(\mu \in \mathbb{T}\) under a secret key \(\mathbf{s}\) as a pair \((\mathbf{a}, b) \in \mathbb{T}^{n+1}\) satisfying \(b = \langle \mathbf{a}, \mathbf{s}\rangle + \mu + e\), where \(e\) denotes a small noise term sampled from the scheme's error distribution.
The \emph{phase} \(\varphi = b - \langle \mathbf{a}, \mathbf{s}\rangle = \mu + e\) thus approximates the original message.
A \emph{TRLWE} ciphertext extends this construction to the polynomial domain, encrypting a message \(\mu(X) \in \mathbb{T}_N[X]\) whose \(N\) coefficients provide up to \(N\) independent plaintext slots under a secret key \(\mathbf{s}(X)\).
Through \emph{sample extraction} and \emph{key switching}, any chosen coefficient of a TRLWE ciphertext can be mapped to a dimension-\(n\) TLWE ciphertext of that coefficient.
A \emph{TRGSW} ciphertext, in turn, encrypts a polynomial and enables the \emph{external product} \(\boxdot : \mathrm{TRGSW} \times \mathrm{TRLWE} \to \mathrm{TRLWE}\), which multiplies the underlying plaintexts while maintaining controlled noise growth.
The external product serves as the core computational primitive for bootstrapping, as detailed in Section~\ref{subsec:pbs}.
A TLWE/TRLWE ciphertext can also be converted into a TRGSW through a costly \emph{circuit bootstrapping} procedure, supporting chained multiplications.

\subsection{Homomorphic Operations}\label{subsec:hom-ops}
Ciphertexts in TFHE are additively homomorphic.
The coordinate-wise sum of two ciphertexts encrypts the sum of their messages.
A ciphertext can also be added to, or scaled by, a cleartext integer constant.
These operations require no bootstrapping, incurring cost linear in the ciphertext size that is negligible relative to a single bootstrap.
Multiplying a TRLWE ciphertext by a cleartext polynomial is likewise bootstrap-free.
However, unlike addition, this operation increases noise proportionally to the norm of the cleartext polynomial's coefficients.

\subsection{Programmable Bootstrapping (PBS)}\label{subsec:pbs}
The defining feature of TFHE is the \emph{PBS}~\cite{CGGI20}, which simultaneously reduces a ciphertext's noise to a fixed level and evaluates an arbitrary function on its message.
Given a TLWE ciphertext encrypting \(\mu \in \mathbb{Z}_p\) and a \emph{test polynomial} \(v(X) \in \mathbb{T}_N[X]\) that encodes a LUT for a function \(f\), a PBS produces a TLWE ciphertext of \(f(\mu)\) with low noise.
The procedure consists of four steps: (i) \emph{modulus switching}, where the input ciphertext is rescaled so that its phase becomes an integer \(\bar\varphi \in \mathbb{Z}_{2N}\), which will serve as a rotation index; (ii) \emph{blind rotation}, where a bootstrapping key, comprising TRGSW encryptions of the secret-key bits, is used to homomorphically compute, via a sequence of external products, a TRLWE ciphertext encrypting \(X^{-\bar\varphi} v(X)\), i.e., the test polynomial rotated by the encrypted input; (iii) \emph{sample extraction}, where the constant coefficient of the rotated polynomial, namely \(v_{\bar\varphi}\), is extracted as a TLWE ciphertext; and (iv) \emph{key switching}, where the resulting TLWE ciphertext is switched back to the original secret key.

Encoding \(f\) amounts to placing \(f(j)\) in the \(j\)-th coefficient of \(v(X)\).
The plaintext modulus \(p\) partitions the \(2N\) rotation positions into \(p\) buckets of width \(2N/p\).
Because the ring is negacyclic, positions in \([N,2N)\) are encoded with a negated sign and thus alias those in \([0,N)\).
A non-negacyclic LUT must therefore confine its values to one half of the ring, leaving \(p/2\) collision-free buckets.
The bucket width \(2N/p\) provides the margin that absorbs input noise and modulus-switching drift, bounding the range of values a single PBS can resolve.
A PBS is by far the most expensive TFHE primitive.
Reducing the number of bootstraps therefore translates almost directly into reduced inference latency.

\subsection{Decision Trees and Private Evaluation}\label{subsec:dt}
A binary decision tree \(\mathcal{T}\) comprises a set of internal (decision) nodes \(\mathcal{I}\) and a set of leaf nodes \(\mathcal{L}\).
Each internal node stores a feature index \(f\) and a threshold \(\tau\), and applies the test \(x_f \leq \tau\) to the corresponding feature of the query vector \(\mathbf{x}\), directing the flow to the left child if the test succeeds and to the right child otherwise.
Each leaf node stores a class label \(y \in \mathcal{Y}\).
Classification of a query proceeds by traversing the unique RTL path determined by the outcomes of these tests, and the tree outputs the label of the reached leaf.

\emph{Depth convention.} We define the \emph{depth} \(\eta\) of a tree as the number of internal node levels along an RTL path, i.e., the number of edges from the root to any leaf.
Under this convention, a complete binary tree of depth \(\eta\) contains \(\mathcal{L} = 2^{\eta}\) leaves, \(2^{\eta}-1\) internal nodes, and \(2^{\eta-1}\) \emph{leaf-parents} (internal nodes with at least one leaf child).
For an arbitrary tree, these quantities serve as upper bounds.

\emph{Quantization.} Since model threshold values and client's inputs are encoded into the coefficient slots of polynomials in TFHE, values are required to be quantized to fit into the polynomial ring.
Quantization constitutes the sole source of deviation from the original floating-point model, and its effect is entirely independent of the homomorphic evaluation itself.
Recent CKKS designs avoid quantization by comparing bit-cast 32-bit inputs~\cite{PCL26}.
Under XCMP this is infeasible, as it would require \(N \ge 2^b\).
Higher-precision TFHE comparators are the corresponding path for our encoding.

\subsection{Homomorphic Comparison}\label{subsec:cmp}
At each internal node, the protocol must homomorphically determine whether the queried feature does not exceed the node's threshold, yielding an encrypted branch indicator \(\varepsilon \in \{0,1\}\).
We adopt the polynomial evaluation version of XCMP, which encodes each \(b\)-bit operand as a monomial.
The comparison bit is read off as a designated coefficient of the product of these TRLWE encrypted monomials with a fixed comparison polynomial, and the encrypted indicator is obtained via sample extraction.
XCMP represents \(b\)-bit operands monomially and therefore requires the ring degree to satisfy \(N \geq 2^{b}\).

\subsection{Path Evaluation and the TE0 Aggregation}\label{subsec:te0}
Given the per-node branch indicators produced by homomorphic comparison, path evaluation accumulates an encrypted cost along each RTL path, yielding a path-cost sum \(s_\ell\) for every leaf \(\ell\).
The SumPath construction~\cite{MNLK23} assigns each edge a cost in \(\{0,1\}\), zero when the comparison at the parent node routes toward \(\ell\) and one otherwise, so that \(s_\ell\) equals the number of mismatching decisions along the path to \(\ell\).
Consequently, \(s_\ell = 0\) for the single active leaf and \(s_\ell \in \{1,\dots,\eta\}\) for every other leaf of a depth-\(\eta\) tree.
Because each \(s_\ell\) is formed exclusively through ciphertext additions, path evaluation employs only additive homomorphic operations and requires no PBS.

To return a single prediction rather than the vector \(\{s_\ell\}\), the path costs must be resolved on the server into the label of the active leaf.
The TE0 aggregation of~\cite{CKNP24} performs this resolution by applying the zero-test indicator
\begin{equation}\label{eq:te0}
    \theta_{\mathrm{EQZero}}(s,\eta)=\frac{1}{\eta!}\prod_{i=1}^{\eta}(i-s)
\end{equation}
to each path sum, which evaluates to \(1\) when \(s=0\) and to \(0\) for every \(s\in\{1,\dots,\eta\}\), acting as an encrypted indicator of the active leaf.
Letting \(y_\ell\) denote the class label stored at leaf \(\ell\), the encrypted prediction is then formed as the additive grouping
\begin{equation}\label{eq:te0-group}
    \mathsf{ct}_{\mathrm{out}}=\sum_{\ell}\theta_{\mathrm{EQZero}}(s_\ell,\eta)\cdot y_\ell,
\end{equation}
in which exactly one term is non-zero, namely that corresponding to the active leaf.
This server-side resolution enables full non-interactivity and facilitates ensemble aggregation, since the client receives only the single ciphertext \(\mathsf{ct}_{\mathrm{out}}\).

The indicator defined in~\eqref{eq:te0} is a degree-\(\eta\) polynomial in the path cost \(s\), evaluated on an RLWE ciphertext of \(s\).
Since TE0 was introduced in the leveled homomorphic setting, it is evaluated through \(O(\eta)\) ciphertext multiplications of multiplicative depth \(O(\log \eta)\), each followed by relinearization, and applied once per leaf.
The cost of this aggregation is therefore \(O(\eta |\mathcal{L}| (ctmult + relinearization))\), making this stage dominate the inference time.
The specific realization of TE0 against which we benchmark, namely its most efficient instantiation in the TFHE setting, as shown in Algorithm~\ref{alg:baseline}, is described in Section~\ref{subsec:baseline}.

\begin{algorithm}[t]
    \caption{TE0 Private Tree Inference (TFHE)}\label{alg:baseline}
    \begin{algorithmic}[1]
        \Require encrypted query \(\mathsf{x}\); tree \(\mathcal{T}\) with internal nodes \(\{(\mathrm{(0, 1)},\tau_v)\}\) and leaf labels \(\{y_\ell\}\); leaf set \(\mathcal{L}\)
        \Ensure encrypted prediction \(\mathsf{ct}_{\mathrm{out}}\)
        \Statex \textit{Stage 1 --- node comparison}
        \For{each internal node \(v\in\mathcal{I}\)}
            \State \(\varepsilon_{lc}(\mathrm{0,1}) \gets \mathsf{XCMP}(\mathsf{x}[\mathrm{feat}(v)], v)\)
            \State \(\varepsilon_{rc} \gets 1 - \varepsilon_{lc}\)
        \EndFor
        \Statex \textit{Stage 2 --- path evaluation}
        \For{each node \(v\in\mathcal{T}\)}
            \State \(\sigma_v \gets \varepsilon_{v} + \sigma_{par}{v}\)
        \EndFor
        \Statex \textit{Stage 3 --- leaf selection}
        \State \(\mathsf{acc} \gets \mathsf{Enc}(0)\)
        \For{each leaf node \(v\in\mathcal{L}\)}
            \State build \(f_v\) according to the leaf's label \(\{y_\ell\}\)
            \State $\mathsf{acc} \gets \mathsf{acc} + \mathsf{BlindRotate}(\mathsf{ModSwitch}(\sigma_v), f_v)$
        \EndFor
        \State $\mathsf{ct}_{\mathrm{out}} \gets \mathsf{KeySwitch}(\mathsf{SampleExtract}(\mathsf{acc}))$
        \State \Return $\mathsf{ct}_{\mathrm{out}}$
    \end{algorithmic}
\end{algorithm}

%% file: method.tex
\subsection{Threat model}
In the non-interactive single-tree private decision-tree evaluation (PDTE) model considered in this work, the client quantizes and encrypts its feature vector under its own secret key and transmits it to the server in a single message.
The server, assumed semi-honest and holding the decision tree in cleartext, evaluates \(\mathcal{T}\) homomorphically over the encrypted query and returns a single ciphertext response.
The client decrypts this ciphertext to obtain the predicted label.
Under this model, client privacy reduces to the IND-CPA security of TFHE. The server learns nothing about the query.
Beyond the prediction, the client may learn the quantization precision b, and we do not claim circuit privacy.
For the gradient-boosted extension, the server returns per-class scores, which reveal the class margins on a per-query basis and, in principle, permit functionally equivalent model extraction under repeated querying~\cite{TZJR+16}.
Both are acceptable under our threat model, which does not treat the model as confidential.
We note, however, that a server-side \(\arg\max\) would reduce this surface for deployments that additionally require model privacy.

\subsection{Overview}\label{subsec:method-overview}
Our construction targets the leaf-selection stage, which we identified as the dominant cost of non-interactive PDTE. Our key observation is that the two leaves sharing a common parent are separated by a single test performed at that parent, so resolving them with two independent per-leaf zero-tests constitutes redundant work.
We instead assign each tree level a signed integer label and pack the two children of each leaf-parent into a single test polynomial.
We then accumulate encrypted root-to-leaf-parent (RTLP) path sums instead of RTL path sums, so that one PBS simultaneously determines which child lies on the active path and returns its class label.
This reduces the PBS count from the number of leaves to the number of leaf-parents.
The encoding that makes this packing well-defined is, as we prove, range-minimal among all valid encodings, whose small label magnitudes additionally reduce the noise incurred during comparison.

The complete inference procedure is given in Algorithm~\ref{alg:inference}: the first two stages, node comparison and path evaluation, are slightly lighter than the baseline (Section~\ref{subsec:te0}) due to our encoding advantage, while the third stage is detailed in Algorithm~\ref{alg:packed-ls}.

\begin{algorithm}[t]
    \caption{Our Private Tree Inference}\label{alg:inference}
    \begin{algorithmic}[1]
        \Require encrypted query \(\mathsf{x}\); tree \(\mathcal{T}\) with internal nodes \(\{(\mathrm{(a_v,b_v)},\tau_v)\}\) and leaf labels \(\{y_\ell\}\); label system \(\mathcal{S}^\ast\); leaf-parent set \(\mathcal{P}\)
        \Ensure encrypted prediction \(\mathsf{ct}_{\mathrm{out}}\)
        \Statex \textit{Stage 1 --- node comparison}
        \For{each internal node \(v\in\mathcal{I}\)}
            \State \(\varepsilon_v(\mathrm{a_v,b_v}) \gets \mathsf{XCMP}(\mathsf{x}[\mathrm{feat}(v)], v)\)
        \EndFor
        \Statex \textit{Stage 2 --- path evaluation}
        \For{each internal node \(v\in\mathcal{I}\)}
            \State \(\sigma_v \gets \varepsilon_v + \sigma_{par}{v}\)
        \EndFor
        \Statex \textit{Stage 3 --- packed leaf selection (Algorithm~\ref{alg:packed-ls})}
        \State \(\mathsf{ct}_{\mathrm{out}} \gets \textsc{PackedLeafSelect}\left(\{\sigma_v\}_{v\in\mathcal{P}}, \mathcal{S}^\ast, \mathcal{P}, \{y_\ell\}\right)\)
        \State \Return \(\mathsf{ct}_{\mathrm{out}}\)
    \end{algorithmic}
\end{algorithm}

\begin{algorithm}[t]
    \caption{Packed Leaf Selection}\label{alg:packed-ls}
    \begin{algorithmic}[1]
        \Require path sums \(\{\sigma_v\}_{v\in\mathcal{P}}\), label system \(\mathcal{S}^\ast\), leaf-parent set \(\mathcal{P}\), class labels \(\{y_\ell\}\)
        \Ensure encrypted prediction \(\mathsf{ct}_{\mathrm{out}}\)
        \State \(\mathsf{acc} \gets \mathsf{Enc}(0)\)
        \For{each leaf-parent \(v\in\mathcal{P}\)}
            \State build \(f_v\) aligning the two children's labels to the reference offsets
            \State $\mathsf{acc} \gets \mathsf{acc} + \mathsf{BlindRotate}(\mathsf{ModSwitch}(\sigma_v), f_v)$
        \EndFor
        \State $\mathsf{ct}_{\mathrm{out}} \gets \mathsf{KeySwitch}(\mathsf{SampleExtract}(\mathsf{acc}))$
        \State \Return \(\mathsf{ct}_{\mathrm{out}}\)
    \end{algorithmic}
\end{algorithm}

\subsection{Signed Dyadic Path Encoding}\label{subsec:encoding}
We encode the tree levels with an integer label system \(\mathcal{S}=\left((a_1,b_1),\dots,(a_\eta,b_\eta)\right)\) in which the node at level \(d\) evaluates to \(a_d\) when the encrypted comparison routes the query to its left child and \(b_d\) otherwise.
Writing \(\varepsilon_d\in\{0,1\}\) for the encrypted branch indicator at level \(d\), with \(\varepsilon_d=0\) for left and \(\varepsilon_d=1\) for right, the contribution of a node on level \(d\) is \((1-\varepsilon_d)a_d+\varepsilon_d b_d\).
The cumulative path sum to any node \(v\) is the prefix sum \(\sigma_v=\sum_{d}\left[(1-\varepsilon_d)a_d+\varepsilon_d b_d\right]\) taken over the ancestors of \(v\), and because every term is selected by a comparison outcome already held as a ciphertext, \(\sigma_v\) is computed by additive homomorphic accumulation alone.
We use the half-symmetric signed dyadic system \(\mathcal{S}^\ast\), in which the root is labeled \((a_1,b_1)=(-1,0)\) and each node at level \(d\ge 2\) is labeled \((a_d,b_d)=(-2^{d-2},+2^{d-2})\).
The corresponding increments \(\delta_d=b_d-a_d\) are \(\delta_1=1\) and \(\delta_d=2^{d-1}\) for \(d\ge2\), i.e.\ the dyadic sequence \(\{1,2,4,\dots,2^{\eta-1}\}\), and the attainable path sums occupy the consecutive range \(\{-2^{\eta-1},\dots,2^{\eta-1}-1\}\).
The motivation for this particular system is established in Section~\ref{subsec:optimal-label}, where it is shown to minimize the path-sum range among all valid encodings.

\subsection{Packed Leaf Selection}\label{subsec:packed-ls}
The leaf-selection stage takes the RTLP path sums \(\sigma_v\) produced by path evaluation (Stage~2 of Algorithm~\ref{alg:inference}) and processes the tree one leaf-parent at a time.
For each leaf-parent \(v\in\mathcal{P}\), we form a single test polynomial \(f_v\) that places the class label of both leaves at the LUT bucket indexed by the path sum of this branch through \(v\), and is zero in every other coefficient.
Concretely, a path-sum value \(s\) indexes the bucket of width \(2N/p\) centered at \(-s\cdot2N/p\) (\(s\leq0\)) or \((p/2-s)\cdot2N/p\) (\(s > 0\)) in the test polynomial, with the class label, encoded as a signed integer centered on zero, written into that bucket.
The bucket width supplies the margin that absorbs the noise in \(\sigma_v\).
When the actual evaluation outcomes place \(v\) on the active path, the path sum coincides with the selected leaf's bucket index, and the PBS returns the label of whichever child the final comparison at \(v\) selects.
When \(v\) does not lie on the active path, at least one ancestor comparison disagrees with this branch, the path sum is displaced from the reference, and the extracted coefficient is zero.

\begin{figure*}[t]
    \centering
    \includegraphics[width=\textwidth]{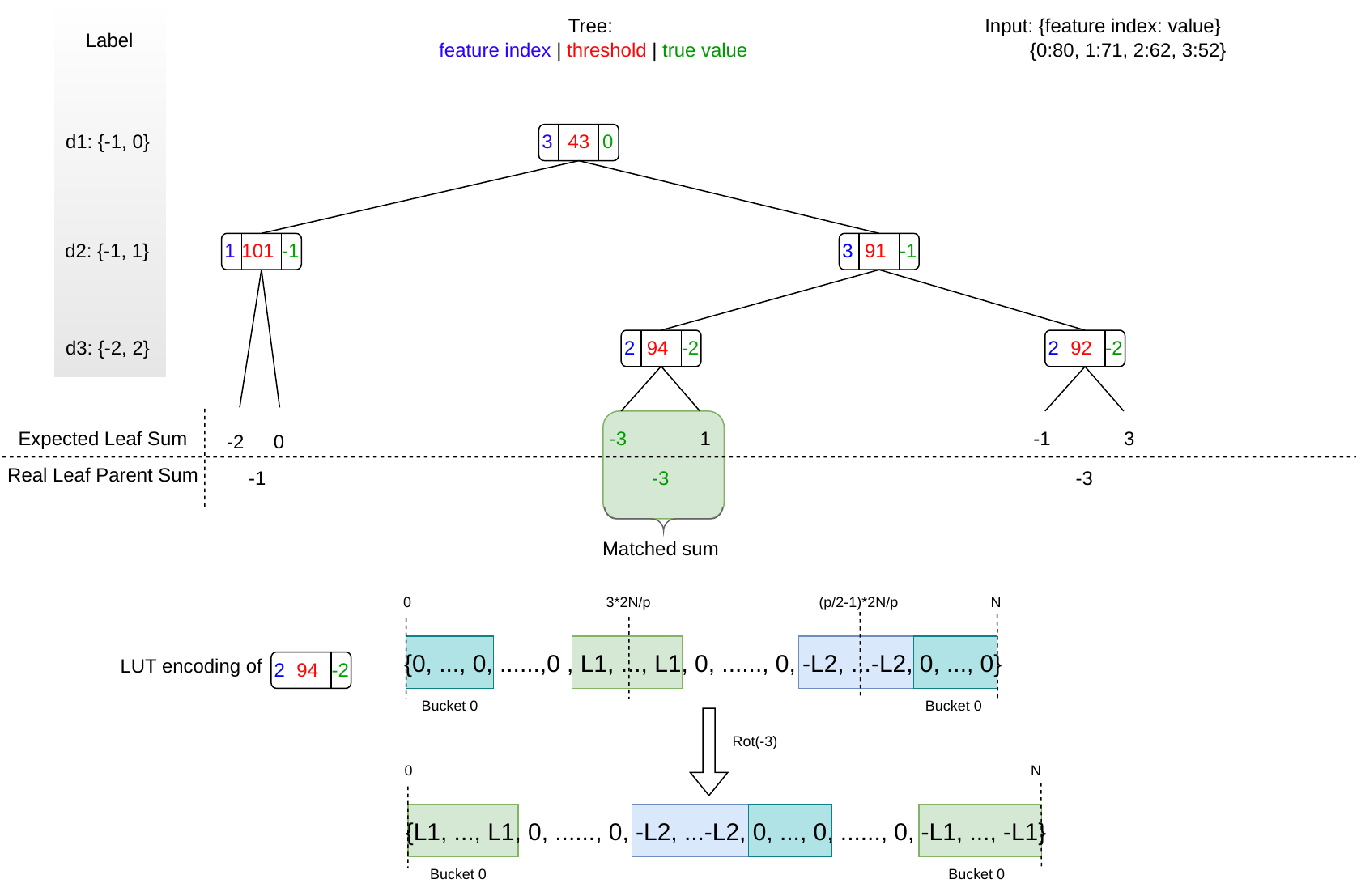}
    \caption{Example of our tree traversal on a depth-3 tree. Each node displays its feature index (blue), threshold (red), and comparison outcome (green) under the input \(\{0{:}80, 1{:}71, 2{:}62, 3{:}52\}\). Per-level dyadic increments (\(d_1\)--\(d_3\)) yield each leaf's Expected Leaf Sum. We pack both children of a leaf-parent into a single PBS LUT, and the leaf whose sum matches the Real Leaf Parent Sum is selected (highlighted, \(-3\)). Below: the LUT encoding of leaf-parent node \((2,94,-2)\), where a blind rotation \(\text{Rot}(-3)\) by the path sum brings the selected label to the constant term.}
    \label{fig:tree}
\end{figure*}

Because at most one leaf-parent lies on the active path, at most one of these per-leaf-parent PBSs returns a non-zero value.
The final prediction is obtained by summing the \(|\mathcal{P}|\) blind-rotation outputs in TRLWE, followed by a single sample extraction and key switch.
The procedure is summarized in Algorithm~\ref{alg:packed-ls}.
Crucially, since each leaf-parent is resolved by its own test polynomial, the construction is entirely local and two distinct leaf-parents may carry identical path sums without affecting correctness, as shown in Figure~\ref{fig:tree}.

\subsection{The Distinct-Sum Property}\label{subsec:dsp}
To make our encoding work, without mismatching different levels, we require the distinct branch configurations along a path to map to distinct sums.
We formalize this as a property of the label system itself, independent of the shape of any particular tree.

\begin{definition}[Integer Binary-Valued Sum System]\label{def:ibvss}
    An \emph{integer binary-valued sum system} (IBVSS) of order \(m\) is a tuple \(\mathcal{S}=\left((a_1,b_1),\dots,(a_m,b_m)\right)\) with \(a_i,b_i\in\mathbb{Z}\) and \(a_i<b_i\).
    Its \emph{increment sequence} is \(\delta_i=b_i-a_i>0\), and its \emph{sum map} is \(\Sigma(\boldsymbol{\varepsilon})=\sum_{i=1}^{m}\left[(1-\varepsilon_i)a_i+\varepsilon_i b_i\right]\), where \(\varepsilon_i \in \{0,1\}\).
\end{definition}

\begin{definition}[Distinct-Sum Property]\label{def:dsp}
\(\mathcal{S}\) satisfies the \emph{distinct-sum property} (DSP) if \(\left|\{\Sigma(\boldsymbol{\varepsilon}):\boldsymbol{\varepsilon}\in\{0,1\}^m\}\right|=2^{m}\), i.e.\ the \(2^{m}\) level combinations are pairwise distinct.
\end{definition}

\begin{proposition}[Characterization]\label{prop:dsp-char}
\(\mathcal{S}\) satisfies the DSP if and only if the only solution of \(\sum_{i=1}^{m}c_i \delta_i=0\) with \(c_i\in\{-1,0,1\}\) is \(c_i=0\) for all \(i\).
\end{proposition}
\begin{proof}
    Two configurations collide, \(\Sigma(\boldsymbol{\varepsilon})=\Sigma(\boldsymbol{\varepsilon}')\), exactly when \(\sum_i(\varepsilon_i-\varepsilon_i')\delta_i=0\).
    Setting \(c_i=\varepsilon_i-\varepsilon_i'\in\{-1,0,1\}\) yields the stated condition, and the absence of a nontrivial solution is equivalent to the absence of a collision.
\end{proof}

The DSP guarantees that the path sums at any single leaf-parent are well-separated regardless of where that leaf sits or how deep the tree's other branches run.
Combined with the local per-leaf-parent construction of Section~\ref{subsec:packed-ls}, this makes each \(f_v\) return the active child's label when \(v\) is active and zero otherwise.
The dyadic increments \(\{1,2,\dots,2^{m-1}\}\) of \(\mathcal{S}^\ast\) form a distinct-subset-sum sequence, satisfying Proposition~\ref{prop:dsp-char}.
The same holds for every prefix \(\{1,2,\dots,2^{d-1}\}\), so the property is preserved at leaves of every depth.

\subsection{Optimal Labeling}\label{subsec:optimal-label}
Among all IBVSS satisfying the DSP we now identify the one minimizing the path sum range, hence more leaves can be encoded into the PBS LUT for a fixed plaintext modulus (Section~\ref{subsec:pbs}). We write \(M(\mathcal{S})=\max_{\boldsymbol{\varepsilon}}\lvert\Sigma(\boldsymbol{\varepsilon})\rvert\) for the worst-case absolute path sum.

\begin{theorem}[Optimal non-negative labeling]\label{thm:nonneg}
Let \(\mathcal{S}\) be an IBVSS of order \(m\) with \(a_i=0\) for all \(i\) satisfying the DSP. Then \(\max_{\boldsymbol{\varepsilon}}\Sigma(\boldsymbol{\varepsilon})=\sum_i\delta_i\ge 2^{m}-1\), with equality if and only if \(\delta_i\) is a permutation of \(\{1,2,4,\dots,2^{m-1}\}\), in which case the image is \(\{0,1,\dots,2^{m}-1\}\).
\end{theorem}

\begin{proof}
    The DSP forces \(2^{m}\) distinct non-negative integers in \(\left[0,\sum_i\delta_i\right]\), so \(\sum_i\delta_i\ge 2^{m}-1\).
    The dyadic choice \(\delta_i=2^{i-1}\) gives \(\Sigma(\boldsymbol{\varepsilon})=\sum_i\varepsilon_i 2^{i-1}\), which enumerates \(\{0,\dots,2^{m}-1\}\) bijectively.
    For uniqueness, order \(\delta_{(1)}\le\cdots\le\delta_{(m)}\): the first \(k\) increments produce \(2^{k}\) distinct subset sums within \(\left[0,\sum_{j\le k}\delta_{(j)}\right]\), which can equal \(\{0,\dots,2^{k}-1\}\) only if \(\delta_{(k+1)}\le 2^{k}\), and the range lower bound applied at each \(k\) forces \(\delta_{(k)}=2^{k-1}\) inductively.
\end{proof}

\begin{theorem}[Optimal signed labeling]\label{thm:signed}
Let \(\mathcal{S}\) be an IBVSS of order \(m\) satisfying the DSP. Then \(M(\mathcal{S})\ge 2^{m-1}\), and equality is achieved by the half-symmetric dyadic system \(\mathcal{S}^\ast=\left((-1,0),(-1,1),(-2,2),\dots,(-2^{m-2},2^{m-2})\right)\), whose image is \(\{-2^{m-1},\dots,2^{m-1}-1\}\).
\end{theorem}

\begin{proof}
    The \(2^{m}\) sums are pairwise distinct integers, so any interval \([-M,M]\) containing them has at least \(2^{m}\) elements, yielding \(2M+1\ge 2^{m}\) and, by integrality of \(M\), \(M\ge 2^{m-1}\).
    For \(\mathcal{S}^\ast\) the increments are \(\{1,2,\dots,2^{m-1}\}\), so the DSP holds by Theorem~\ref{thm:nonneg}.
    The offset is \(\sum_i a_i=-1-1-2-\cdots-2^{m-2}=-2^{m-1}\), and the image is the translation \(\{-2^{m-1},\dots,-2^{m-1}+(2^{m}-1)\}=\{-2^{m-1},\dots,2^{m-1}-1\}\), giving \(M(\mathcal{S}^\ast)=2^{m-1}\).
\end{proof}

\begin{figure}[t]
    \centering
    \includegraphics[width=\columnwidth]{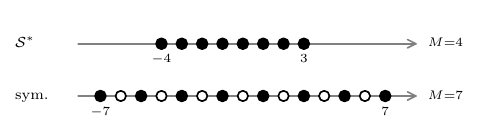}
    \caption{Path-sum images for \(m=3\). The optimal half-symmetric signed dyadic system \(\mathcal{S}^\ast\) (top) fills a gap-free interval of \(2^{m}=8\) consecutive integers with range \(M=2^{m-1}=4\). The full-symmetric labeling \((-2^{m-1},+2^{m-1})\) (bottom) is parity-locked: it occupies only the odd integers and wastes every even slot, spanning twice the range, \(M=7\).}
    \label{fig:numberline}
\end{figure}

Theorem~\ref{thm:nonneg} establishes that the dyadic increment sequence is the unique range-minimal encoding satisfying the DSP, and Theorem~\ref{thm:signed} shows that admitting signed labels reduces the worst-case path sum from \(2^m-1\) to \(2^{m-1}\).
We emphasize that both the signed optimal encoding \(\mathcal{S}^\ast\) and the unsigned dyadic system produce exactly \(2^m\) distinct path sums and therefore occupy the same number of LUT buckets.
Minimizing the range does not extend the maximum supportable tree depth, which is governed by the bucket capacity constraint \(2^{\eta} \le p/2\).
The benefit of the signed encoding over its unsigned counterpart is instead a noise advantage.
The smaller label magnitudes of \(\mathcal{S}^\ast\) reduce the noise introduced when the labels are combined with the encrypted query during comparison, as quantified in Section~\ref{subsubsec:sd-noise}.
The centering of \(\mathcal{S}^\ast\) also distinguishes it from a full-symmetric assignment such as
\((-2^{m-1},+2^{m-1})\), whose increments \(\{2,4,\dots,2^{m}\}\) share a common factor of two and leave every alternate integer unused, thereby doubling the span.
Table~\ref{tab:example} lists \(\mathcal{S}^\ast\) for \(m=3\), and Figure~\ref{fig:numberline} contrasts its gap-free image with the parity-locked image of the symmetric assignment.

\begin{table}[t]
    \centering
    \caption{The optimal signed dyadic system \(\mathcal{S}^\ast\) for \(m=3\). The attainable path sums form the gap-free interval \(\{-4,\dots,3\}\), attaining the optimal range \(M=2^{m-1}=4\).}
    \label{tab:example}
    \begin{tabular}{ccc}
        \toprule
        level \(d\) & \((a_d,b_d)\) & \(\delta_d\) \\
        \midrule
        1 & \((-1,0)\) & 1 \\
        2 & \((-1,1)\) & 2 \\
        3 & \((-2,2)\) & 4 \\
        \bottomrule
    \end{tabular}
\end{table}

\subsection{Correctness and Failure Probability}\label{subsec:noise}
TFHE-based SumPath PDTE is correct except with a per-operation probability determined by the noise at two points of the pipeline.
We state the two conditions under which Algorithm~\ref{alg:inference} returns the exact plaintext prediction.
We also derive the failure probability from the accumulated noise variances, and choose parameters (Tables~\ref{tab:params} and~\ref{tab:runtime}) such that the per-inference failure probability is at most \(2^{-64}\).

\subsubsection{Correctness conditions}
Fix a leaf-parent \(v\in\mathcal{P}\) and let \(\sigma_v\) denote its encrypted path sum after Stage~2.
Write \(e^{\mathrm{in}}_v\) for the phase error of \(\sigma_v\) at the input of the blind rotation, \emph{after} the modulus switch of the PBS. Let \(e^{\mathrm{out}}\) denote the error of the final ciphertext \(\mathrm{ct}_{\mathrm{out}}\) decrypted by the client.
All errors are expressed on the torus, so the half-bucket decryption margin at plaintext modulus \(p\) is \(1/(2p)\).

\begin{lemma}[Packed leaf selection]
    \label{lem:correctness}
    Let \(\mathcal{S}^{*}\) satisfy the DSP, let \(2^{\eta} \le p/2\), and let \(f_v\) be constructed as in Section~\ref{subsec:packed-ls}.
    If
    \begin{equation}
        \left|e^{\mathrm{in}}_v\right| < \frac{1}{2p},
        \label{eq:cond-in}
    \end{equation}
    then \(\mathrm{BlindRotate}(\sigma_v, f_v)\) encrypts the class label of the active child of \(v\) when \(v\) lies on the active path, and encrypts \(0\) otherwise.
    If moreover the aggregated output satisfies
    \begin{equation}
        \left|e^{\mathrm{out}}\right| = \left|\sum_{v\in\mathcal{P}} e^{\mathrm{br}}_v + e^{\mathrm{se+ks}}\right| < \frac{1}{2p},
        \label{eq:cond-out}
    \end{equation}
    where \(e^{\mathrm{br}}_v\) is the fresh output error of the \(v\)-th blind rotate and \(e^{\mathrm{se+ks}}\) is the final sample extract and key switch error, then the client decrypts the exact prediction of the plaintext tree.
\end{lemma}

\begin{proof}
    By the DSP, the \(2^{d}\) branch configurations at depth \(d\) map to pairwise distinct path sums, and by \(2^{\eta} \le p/2\) these sums index pairwise distinct, non-aliasing LUT buckets.
    The polynomial \(f_v\) places the two children's labels in the two buckets indexed by the path sums that agree with every ancestor comparison of \(v\), and zero elsewhere.
    Any configuration in which at least one ancestor comparison disagrees selects a zero bucket.
    Under condition~\eqref{eq:cond-in} the rotation index lands strictly inside the selected bucket, so the extracted coefficient equals the active child's label if \(v\) is on the active path, \(0\) otherwise.
    Since exactly one leaf-parent lies on the active path, the sum of the \(|\mathcal{P}|\) outputs encrypts the active leaf's label with error \(e^{\mathrm{out}}\).
\end{proof}

\subsubsection{Noise budget}
We now bound the variances of \(e^{\mathrm{in}}_v\) and \(e^{\mathrm{out}}\) from the pipeline of Algorithm~\ref{alg:inference}.
Let \(\hat\sigma^{2}_{\mathrm{lwe}}\) and \(\hat\sigma^{2}_{\mathrm{rlwe}}\) denote the torus-normalized variances of the error distributions.
We compute the variances under the standard independence heuristic, with balanced digits of uniform values and binary keys at average weight.

\paragraph{Input condition}
The error of \(\sigma_v\) comprises three independent contributions:
\begin{equation}
    \mathrm{Var}\left(e^{\mathrm{in}}_v\right) \leq V_{\mathrm{cmp}} + V_{\mathrm{se+ks}} + V_{\mathrm{ms}}.
    \label{eq:var-in}
\end{equation}

\emph{(i) Comparison.} At level \(d\) the label enters as the coefficients of a cleartext polynomial multiplied with the encrypted query.
The polynomial is filled with \(\beta=2^b\) coefficients of magnitude \(|a_d|\vee|b_d|\), and an attribute may be tested several times, so
\begin{equation}
    V_{\mathrm{cmp}} < 3 \sum_{d=1}^{\eta} \max(a_d^{2}, b_d^{2}) \beta \hat\sigma^{2}_{\mathrm{rlwe}}.
    \label{eq:var-cmp}
\end{equation}

\emph{(ii) Sample extraction and key switching}, performed once per leaf-parent, with an exact gadget \(\ell_{\mathrm{ks}} \log_2 B_{\mathrm{ks}} = \log_2 q_{\mathrm{lwe}}\):
\begin{equation}
    \begin{split}
    V_{\mathrm{se+ks}} &= \left( 1 + \frac{kN}{2} \right) \frac{q_{\mathrm{lwe}}^{-2} - q_{\mathrm{rlwe}}^{-2}}{12} \\
    &\qquad + kN \ell_{\mathrm{ks}} \frac{B_{\mathrm{ks}}^2 + 2}{12} \hat{\sigma}_{\mathrm{lwe}}^2,
    \end{split}
    \label{eq:var-ks}
\end{equation}
where the first term is the rounding noise of the \(q_{\mathrm{rlwe}} \rightarrow q_{\mathrm{lwe}}\) performed before every key switch.

\emph{(iii) Modulus switching} inside the PBS, for a binary secret of dimension \(n\):
\begin{equation}
    V_{\mathrm{ms}} = \frac{1 + n/2}{48N^{2}}.
    \label{eq:var-ms}
\end{equation}

\paragraph{Output condition}
Each PBS output carries fresh blind-rotation noise.
With \(\ell_{\mathrm{a}}\) decomposition levels used out of \(\ell\),
\begin{equation}
    \begin{split}
        V_{\mathrm{br}}
        &= n\left[(k+1)\ell_{\mathrm{a}} N \frac{B^{2}+2}{12} \hat{\sigma}^{2}_{\mathrm{rlwe}} \right.\\
        &\qquad \left. + \left(1+\frac{kN}{2}\right)\frac{B^{-2\ell_{\mathrm{a}}} - q_{\mathrm{rlwe}}^{-2}}{12}\right],
    \end{split}
    \label{eq:var-br}
\end{equation}
where the second term is the truncation residual of the approximate decomposition.
The final error variance is \(\mathrm{Var}(e^{\mathrm{out}}) \le |\mathcal{P}| V_{\mathrm{br}} + V_{\mathrm{se+ks}}\).

\subsubsection{Failure probability}
Both errors are sums of many independent contributions, and we model them as Gaussian.
Writing \(m = 1/(2p)\) and \(\sigma\) for the standard deviation of the relevant error, a single condition fails with probability \(\operatorname{erfc}\!\left(m/\left(\sigma\sqrt{2}\right)\right)\).
Condition~\eqref{eq:cond-in} must hold at each of the \(|\mathcal{P}|\) blind rotations and condition~\eqref{eq:cond-out} once, so by a union bound the per-inference failure probability is
\begin{equation}
    p_{\mathrm{error}} \le |\mathcal{P}| \operatorname{erfc}\!\left(\tfrac{m}{\sigma_{\mathrm{in}}\sqrt{2}}\right)
    + \operatorname{erfc}\!\left(\tfrac{m}{\sigma_{\mathrm{out}}\sqrt{2}}\right).
    \label{eq:pfail}
\end{equation}

\subsubsection{The signed encoding as noise margin}\label{subsubsec:sd-noise}
Under the unsigned dyadic labeling, the label magnitude at level \(d\) is \(2^{d-1}\).
Under \(\mathcal{S}^*\), it is 1 at the root and \(2^{d-2}\) for \(d \ge 2\), so it is half as large at every level below the root.
The sum \(\sum_d \max(a_d^2,b_d^2)\) in Eq.~\eqref{eq:var-cmp} therefore drops from \((4^\eta-1)/3\) to \((4^{\eta-1}+2)/3\).
This is a factor of 2.5, 3.5, and 3.9 at \(\eta=2,3,4\), and the factor approaches 4 as \(\eta\) grows.
At the parameters we use (Tables~\ref{tab:params} and~\ref{tab:runtime}) this advantage manifests as additional margin, since \(V_{\mathrm{cmp}}\) is dominated by \(V_{\mathrm{ms}}\) by several orders of magnitude.
It becomes material in regimes where the comparison noise is dominant, e.g.\ larger relative RLWE noise or substantially deeper trees.

\subsection{Support for Arbitrary Trees}\label{subsec:arbitrary}
Because the DSP is a property of the level encoding and the leaf-selection construction is local to each leaf-parent, the method applies to trees of arbitrary shape without padding them to a complete form.
A leaf-parent with two leaf children is resolved by a single PBS as described, while an internal node with one leaf child and one internal child contributes that single leaf through one PBS in the same manner.
The number of PBSs is the number of leaf-parents, which ranges from \(|\mathcal{L}|/2\) for a complete or balanced tree, up to \(|\mathcal{L}|\) where no two leaves share a parent.
Consequently, the PBS saving over the per-leaf TE0 baseline lies between \(1\times\) and \(2\times\), reaching the full \(2\times\) on balanced trees, without introducing dummy nodes.

\subsection{Cost Analysis}\label{subsec:cost}
The node-comparison and path-sum stages are realized by homomorphic comparisons and additive accumulation, whose cost is much smaller than PBS. The leaf-selection stage dominates the overall cost in both schemes.
Nonetheless, our encoding makes the first two stages slightly lighter than the baseline.
In node comparison the baseline TE0 forms each internal node's two branch indicators as \(\varepsilon_{lc}\gets\mathsf{XCMP}(\cdot)\) and its complement \(\varepsilon_{rc}\gets 1-\varepsilon_{lc}\), whereas our signed encoding represents the branch by the single indicator \(\varepsilon_v\) directly returned by \(\mathsf{XCMP}\), eliminating the per-node complement.
In path summation the baseline accumulates every node's value into all \(|\mathcal{L}|\) leaves, while we accumulate only to the \(|\mathcal{P}|\) leaf-parents.

The optimal TFHE realization of the baseline performs \(|\mathcal{L}|\) PBSs, one per leaf, each evaluating a LUT that maps the zero path sum to the corresponding leaf value and every other sum to zero.
The leaf-values are encoded into each leaf's LUT, making leaf label multiplication absorbed by PBS. While our packed leaf selection performs one PBS per leaf-parent.
The response returned to the client remains a single TLWE ciphertext, identical in size to the baseline, and the entire evaluation is non-interactive.
Thus, our method preserves the communication profile and the correctness of the baseline while reducing the PBS work from \(|\mathcal{L}|\) to \(|\mathcal{P}|\).

Our encoding assigns every RTL configuration to a distinct LUT bucket, thereby consuming \(2^{\eta}\) of the \(p/2\) collision-free buckets, whereas the baseline distinguishes only the \(\eta+1\) mismatch counts.
This exponential-versus-linear bucket demand constitutes the trade-off incurred by reducing the PBS workload.
Consequently, the baseline is able to employ more lightweight parameters than ours, yielding a lower per-PBS cost.
To thoroughly examine this trade-off, we derive cost-optimal parameters for both schemes over the range \(\eta \in [1,5]\), with the detailed numerical results provided in the supplementary material.

Under the at least \(128\)-bit security and \(p_{\mathrm{error}} \leq 2^{-64}\) correctness conditions, the cost model predicts our method ahead of TE0 at every depth \(1\) through \(4\), with the margin contracting as depth grows.
At depth \(5\), our \(N_{\min}\) crosses \(2048\).
The doubled ring degree offsets the halved PBS count, dropping the ratio below parity.
We read the model for the ordering it induces and the crossover depth it locates: the measured latencies of Section~\ref{subsec:latency} follow this ranking but run below the model's estimate, as it omits the equal-cost leaf-grouping stage and the fixed per-operation overheads that dilute the end-to-end saving.
Our profitable region is therefore \(\eta \leq 4\).
This region coincides with the trees that dominate practical tabular learning: gradient-boosted ensembles use trees of depth \(3\)--\(8\)~\cite{BB19}, typically \(\le 4\), and every tree of such an ensemble enjoys the leaf-selection reduction independently (Section~\ref{subsec:ensemble}).
For substantially deeper trees, the baseline's linear bucket demand prevails, and when many queries can be batched, square-root-scaling designs such as~\cite{PCL26} become preferable.

%% file: experiment.tex
\subsection{Implementation and Setup}\label{subsec:setup}
We implement both our scheme and the TE0 baseline in C++ atop an in-house TFHE library, with shared low-level kernels (NTT, polynomial arithmetic, external product, key-switching)\footnote{https://github.com/xsong93/sdpe-pdte/tree/main}.
We grant each scheme its individually cost-optimal configuration \((p, n, N, \ell_a)\) at every depth (selected under identical constraints of at least \(128\)-bit security and \(p_{\mathrm{error}} \le 2^{-64}\), reported in the note to Table~\ref{tab:runtime}). The comparison isolates the encoding under shared parameters where the two schemes' plaintext ranges coincide.
It lets the baseline descend to the lighter dimension its smaller range affords where they diverge, pricing the full system-level cost of adopting our encoding.
The single-tree experiments use a \(32\)-bit RLWE modulus, which halves the per-attribute query relative to the \(56\)-bit torus and keeps the uplink at the minimum the setting assumes.
The ensemble requires the wider torus for the reason given in Section~\ref{subsec:discuss-param-gbdt}.

All homomorphic operations are instantiated at the \(128\)-bit security level, with the cryptographic parameters listed in Table~\ref{tab:params}.
The plaintext modulus varies with depth and scheme.
Single-tree class labels are encoded directly within the range and centered on zero, e.g., the seven Steel classes map to \(\{-3,\dots,-1,1,\dots,4\}\).
Because an off-path leaf-parent extracts zero, the value zero is reserved as the marker for an empty bucket and is never used as a class label.
For the multi-class gradient-boosted ensemble, the per-class accumulated scores are quantized to the interval \([-2^{\kappa-1}, 2^{\kappa-1})\), hence leaf weights are encoded on a torus grid of step \(2^{-\kappa} \cdot q\).
Decision trees were trained with scikit-learn on real-valued features, then quantized to \(b=6\)-bit integers together with their thresholds prior to encryption.
All experiments were run on an AMD Ryzen~9 9950X3D CPU with \(128\) GB RAM under Ubuntu~24.04.4 LTS.

All experiments exploit parallelism at the granularity of individual trees.
Single-tree inference runs on a single thread, while the gradient-boosted ensemble evaluation distributes the constituent trees across \(16\) threads, corresponding to the maximum core count of our CPU.

\begin{table}[t]
    \centering
    \begin{threeparttable}
        \caption{Cryptographic parameters. Single-tree configurations are cost-optimal per depth and scheme and are listed in Table~\ref{tab:runtime}; the values below are shared across all experiments (top) or specific to the gradient-boosted ensemble (bottom).}
        \label{tab:params}
        \begin{tabular}{ll}
            \toprule
            Parameter & Value \\
            \midrule
            \multicolumn{2}{l}{\textit{Common to all experiments}} \\
            secret distribution                                & binary \\
            LWE ciphertext modulus \(q_{\mathrm{lwe}}\)        & \(2^{32}\) \\
            key switch GD base / \(\ell\)                      & \((2^{2}, 16)\) \tnote{3} \\
            node quantization bit \(b\)                        & \(6\) \\
            failure probability \(p_{\mathrm{error}}\)         & \(\leq 2^{-64}\) per inference \\
            \midrule
            \multicolumn{2}{l}{\textit{Gradient-boosted ensemble}} \\
            RLWE ciphertext modulus \(q_{\mathrm{rlwe}}\)      & \(2^{56}\) \\
            ring degree \(N\)                                  & \(2048\) \\
            LWE dimension \(n\)                                & \(1024\) \\
            LWE error \(\chi_{\mathrm{lwe}}\)                  & \(\mathrm{TUniform}(10)\), \(\log_2(\sigma) \approx 9.21\) \tnote{2} \\
            RLWE error \(\chi_{\mathrm{rlwe}}\)                & \(\mathrm{TUniform}(8)\), \(\log_2(\sigma) \approx 7.21\) \tnote{2} \\
            TRGSW GD base / \(\ell\) / \(\ell_a\)              & \((2^{14}, 4, 2)\) \tnote{4} \\
            score quantization bit \(\kappa\)                  & \(10\) \\
            \bottomrule
        \end{tabular}
        \begin{tablenotes}\footnotesize
        \item[1] All parameter sets provide \(\geq 128\)-bit security for both the LWE and RLWE instances, estimated with the Lattice Estimator~\cite{APS15} (commit~\texttt{3e48ef4}, MATZOV cost model), minimizing over primal, dual, and hybrid attacks.
        \item[2] Error distributions follow Table 5.3 of~\cite{BCCC+25}; single-tree noise bounds are per configuration (Table~\ref{tab:runtime}).
        \item[3] KS uses the LWE parameter set, operating on a 32-bit torus, with full decompose gadget.
        \item[4] Ciphertext-plaintext multiplication uses \(\ell\); the blind rotation uses \(\ell_a\).
        \end{tablenotes}
    \end{threeparttable}
\end{table}

\subsection{Baseline}\label{subsec:baseline}
As discussed in Section~\ref{subsubsec:gap}, the state-of-the-art for resolving the active leaf on the server is the TE0 aggregation of~\cite{CKNP24}.
We therefore adopt TE0 as our baseline.
Because TE0 was introduced in the leveled homomorphic setting and no TFHE realization of it exists in the literature, a fair comparison in the TFHE regime obliges us to instantiate it on TFHE ourselves.
We implement it in its most efficient form, evaluating each per-leaf zero-test as a single PBS whose LUT maps the zero path sum directly to that leaf's class label and every other sum to zero.
Folding the leaf-value multiplication into the PBS lets the baseline perform exactly \(|\mathcal{L}|\) PBSs and no follow up leaf-value products.
This TFHE instantiation of TE0 is strictly cheaper than evaluating Eq.~\ref{eq:te0} as a polynomial-evaluation circuit under TFHE, so every speedup we report is measured against the most efficient realization of the per-leaf TE0 zero test.
Both schemes are evaluated on identical tree structures, and neither applies leaf-level pruning, as discussed in Section~\ref{subsec:discuss-trunc}.
The comparison therefore isolates the effect of the encoding.

\subsection{Datasets}\label{subsec:datasets}
We evaluate on four UCI classification datasets widely used in prior PDTE work~\cite{MNLK23,CKNP24}: Breast Cancer (Cancer), Heart Disease (Heart), Spambase, and Steel Plates Faults (Steel), summarized in Table~\ref{tab:datasets}.
Cancer, Heart, Spambase are binary, whereas Steel (\(7\) classes) is multi-class, so the evaluation exercises our leaf encoding across both regimes.
For each dataset, we train trees of depths \(1\)--\(4\).

\begin{table}[t]
    \centering
    \caption{Datasets used in the evaluation.}
    \label{tab:datasets}
    \begin{tabular}{lccc}
        \toprule
        Dataset & Features & Instances & Classes \\
        \midrule
        Cancer   & 30 & 569  & 2 \\
        Heart    & 13 & 303  & 2 \\
        Spambase & 57 & 4601 & 2 \\
        Steel    & 27 & 1941 & 7 \\
        \bottomrule
    \end{tabular}
\end{table}

\subsection{Effect of Quantization Precision}\label{subsec:precision}
\begin{figure}[t]
    \centering
    \includegraphics[width=\columnwidth]{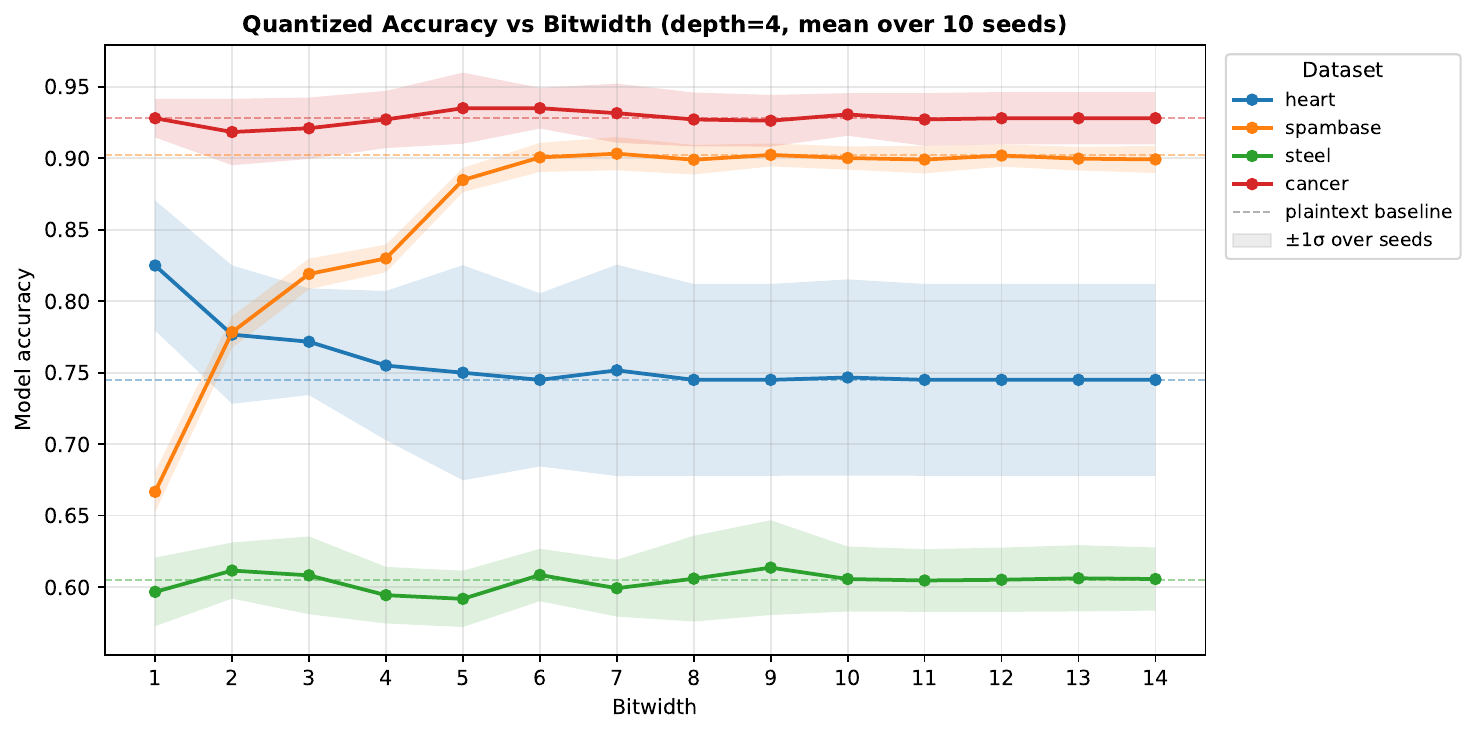}
    \caption{Model accuracy versus bitwidth at depth~\(4\). Dashed lines mark each dataset's floating-point baseline. At \(b=6\) the model accuracy matches the baseline on every dataset.}
    \label{fig:bitwidth}
\end{figure}

Because features and thresholds are quantized to \(b\)-bit integers, the precision \(b\) governs how faithfully the quantized model reproduces the original floating-point tree.
We study accuracy as a function of \(b\) to identify the smallest precision that preserves the model's behavior.
Figure~\ref{fig:bitwidth} reports the quantized-model accuracy at depth~\(4\) for bitwidths from \(1\) to \(14\), with each dataset's floating-point accuracy shown as a dashed baseline.

We adopt \(b=6\) for all remaining experiments for two reasons.
First, it is sufficient since quantized accuracy has converged to the floating-point baseline across datasets, so no information is lost.
Second, it is economical since precision affects polynomial ring size, \(b=6\) fits perfectly inside our \(11\)-bit ring.

\subsection{Correctness}\label{subsec:correctness}
Because both schemes use exact integer arithmetic, the encrypted inference reproduces the quantized-plaintext prediction without approximation.
Across all datasets and depths, the encrypted output of our scheme agreed with the quantized-plaintext model on every test instance (consistent with the \(\leq 2^{-64}\) bound), and matched the baseline output exactly.
The two schemes are functionally identical and differ only in cost.

\subsection{Latency Analysis and Component Breakdown}\label{subsec:latency}
Table~\ref{tab:runtime} reports the end-to-end single-tree inference latency of our scheme against the baseline across the four datasets and depths \(1\)--\(4\), with each \((\text{scheme}, \eta)\) pair benchmarked at its own cost-optimal parameter set.
Our method is faster at every depth, by \(\SpeedupLo\times\)--\(\SpeedupHi\times\).
The dominant driver is the PBS-count saving \(R = |\mathcal{L}| / |\mathcal{P}|\): \(R = 2\) for depths \(1\)--\(3\) and \(1.7\)--\(2\) at depth \(4\), realizing the reduction predicted by Table~\ref{tab:complexity}.
Because the class-label range floors the plaintext modulus at \(p=16\) for both schemes at depths \(1\)--\(3\), the two pipelines run at identical parameters there and the measured ratio isolates the encoding, staying a uniform \(1.8\times\)--\(1.9\times\) across these depths.
The marginally lower ratio at depth \(1\) is a dilution effect.
The fixed-cost leaf-grouping stage, common to both schemes, is proportionally largest when the per-tree PBS work is smallest, and its share shrinks as depth grows.
Only at depth \(4\) does our path-sum demand force \(p=32\) while the baseline retains \(p=16\), so depth \(4\) prices the full system-level swap, with the baseline granted the lighter configuration its smaller range affords, narrowing our advantage there to \(1.5\times\)--\(1.8\times\).

To locate the source of this speedup, Table~\ref{tab:runtime} further decomposes the runtime into the leaf-selection PBS and the remaining comparison, path-evaluation, and leaf-grouping operations.
Leaf selection accounts for the vast majority of the cost in both schemes, and its reduction tracks the overall trend.
It is near the full halving where the two schemes share parameters at depths \(1\)--\(3\), and narrows at depth \(4\) as the baseline's lighter configuration offsets part of the halved PBS count.
The path-sum stage is likewise lighter under our encoding, since the per-node complement is eliminated and accumulation targets only the \(|\mathcal{P}|\) leaf-parents.
Leaf grouping, by contrast, acts on the single aggregated output and costs the same in both, diluting the end-to-end ratio slightly below the leaf-selection ratio.
Being small relative to leaf selection, these stages move the overall figure only marginally.

\begin{table*}[t]
    \centering\footnotesize
    \begin{threeparttable}
        \caption{Per-stage decision-tree inference CPU time, one query at a time, mean over 50 queries\tnote{2}, \(b=6\)}
        \label{tab:runtime}
        \begin{tabular}{l c r r r r r r r r r r c}
            \toprule
            Dataset & \(\eta\) & \multicolumn{5}{c}{Ours (ms)} & \multicolumn{5}{c}{TE0 (ms)} & speedup\tnote{1} \\
            \cmidrule(lr){3-7} \cmidrule(lr){8-12}
             &  & cmp & sum & sel & grp & total & cmp & sum & sel & grp & total &  \\
            \midrule
            \multirow{4}{*}{Cancer} & 1 & 0.0 & 4.0 & 18.0 & 4.1 & 26.2 & 0.0 & 7.4 & 36.7 & 4.1 & 48.3 & \(\mathbf{1.84\times}\) \\
             & 2 & 0.1 & 6.8 & 37.1 & 3.8 & 47.8 & 0.1 & 13.3 & 73.4 & 4.0 & 90.8 & \(\mathbf{1.90\times}\) \\
             & 3 & 0.1 & 13.4 & 73.9 & 4.1 & 91.6 & 0.1 & 23.7 & 145.9 & 4.0 & 173.7 & \(\mathbf{1.90\times}\) \\
             & 4 & 0.2 & 24.4 & 274.2 & 4.4 & 303.3 & 0.2 & 34.8 & 454.6 & 3.7 & 493.3 & \(\mathbf{1.63\times}\) \\
            \midrule
            \multirow{4}{*}{Heart} & 1 & 0.0 & 4.0 & 18.3 & 4.2 & 26.4 & 0.0 & 7.3 & 36.2 & 4.1 & 47.7 & \(\mathbf{1.80\times}\) \\
             & 2 & 0.1 & 6.7 & 36.8 & 3.8 & 47.4 & 0.1 & 12.7 & 72.7 & 4.1 & 89.6 & \(\mathbf{1.89\times}\) \\
             & 3 & 0.1 & 13.0 & 73.7 & 4.1 & 91.0 & 0.1 & 23.1 & 145.8 & 4.0 & 173.0 & \(\mathbf{1.90\times}\) \\
             & 4 & 0.1 & 27.2 & 312.3 & 4.5 & 344.1 & 0.2 & 41.2 & 561.2 & 3.7 & 606.3 & \(\mathbf{1.76\times}\) \\
            \midrule
            \multirow{4}{*}{Spambase} & 1 & 0.0 & 4.1 & 18.5 & 4.3 & 26.9 & 0.0 & 7.4 & 36.4 & 4.1 & 48.1 & \(\mathbf{1.79\times}\) \\
             & 2 & 0.1 & 6.8 & 36.7 & 3.8 & 47.3 & 0.1 & 13.0 & 72.8 & 4.2 & 90.1 & \(\mathbf{1.91\times}\) \\
             & 3 & 0.1 & 13.2 & 73.4 & 4.2 & 90.9 & 0.1 & 23.3 & 145.8 & 3.9 & 173.2 & \(\mathbf{1.91\times}\) \\
             & 4 & 0.1 & 27.3 & 311.4 & 4.5 & 343.2 & 0.2 & 37.1 & 490.2 & 3.8 & 531.3 & \(\mathbf{1.55\times}\) \\
            \midrule
            \multirow{4}{*}{Steel} & 1 & 0.0 & 4.0 & 18.3 & 4.1 & 26.5 & 0.0 & 7.4 & 36.4 & 4.1 & 48.0 & \(\mathbf{1.81\times}\) \\
             & 2 & 0.1 & 6.8 & 36.9 & 3.7 & 47.4 & 0.1 & 13.0 & 73.2 & 4.2 & 90.5 & \(\mathbf{1.91\times}\) \\
             & 3 & 0.1 & 13.0 & 73.4 & 4.2 & 90.7 & 0.1 & 23.2 & 146.4 & 4.0 & 173.7 & \(\mathbf{1.92\times}\) \\
             & 4 & 0.1 & 24.2 & 272.7 & 4.4 & 301.6 & 0.1 & 32.2 & 420.3 & 3.7 & 456.3 & \(\mathbf{1.51\times}\) \\
            \bottomrule
        \end{tabular}
        \begin{tablenotes}\footnotesize
        \item[1] Stages: cmp = compare tree, sum = sum path, sel = select leaf, grp = group leaf.
        \item[2] Each (scheme, \(\eta\)) pair is benchmarked at \emph{its own} parameter set: the cost-optimal configuration for that scheme at that depth under \(p_{\mathrm{error}}\le 2^{-64}\) and at least \(128\)-bit security. All single-tree configurations use \(q_{\mathrm{lwe}}=q_{\mathrm{rlwe}}=2^{32}\). Sets used here (\(B=2^{\beta_g}\) the gadget base, \(\ell\) its full length, \(\ell_a\) the levels the blind rotation uses; \(\mathcal{U}_{\mathrm{LWE}}\) and \(\mathcal{U}_{\mathrm{RLWE}}\) the TUniform bound bits \(u\) of the LWE/KSK and RLWE/BSK noise) -- \emph{Ours} \(\eta=1\text{--}3\): \((p,n,N,B,\ell,\ell_a)=(16,760,2048,2^{10},3,2)\), \(\mathcal{U}_{\mathrm{LWE}}=16\), \(\mathcal{U}_{\mathrm{RLWE}}=3\); \(\eta=4\): \((p,n,N,B,\ell,\ell_a)=(32,840,2048,2^{7},4,3)\), \(\mathcal{U}_{\mathrm{LWE}}=14\), \(\mathcal{U}_{\mathrm{RLWE}}=3\). \emph{TE0} \(\eta=1\text{--}3\): \((p,n,N,B,\ell,\ell_a)=(16,760,2048,2^{10},3,2)\), \(\mathcal{U}_{\mathrm{LWE}}=16\),
            \(\mathcal{U}_{\mathrm{RLWE}}=3\); \(\eta=4\): \((p,n,N,B,\ell,\ell_a)=(16,760,2048,2^{7},4,3)\), \(\mathcal{U}_{\mathrm{LWE}}=16\), \(\mathcal{U}_{\mathrm{RLWE}}=3\).
        \end{tablenotes}
    \end{threeparttable}
\end{table*}

\subsection{Communication}\label{subsec:comm}
Both schemes are non-interactive with identical communication profile.
The client uploads a single encrypted query (208-912 KB, varies by dataset feature count) and receives a single TLWE ciphertext (3 KB) encoding the prediction.
The compact single-ciphertext response of the baseline is preserved exactly.

\begin{figure}[t]
    \centering
    \includegraphics[width=\columnwidth]{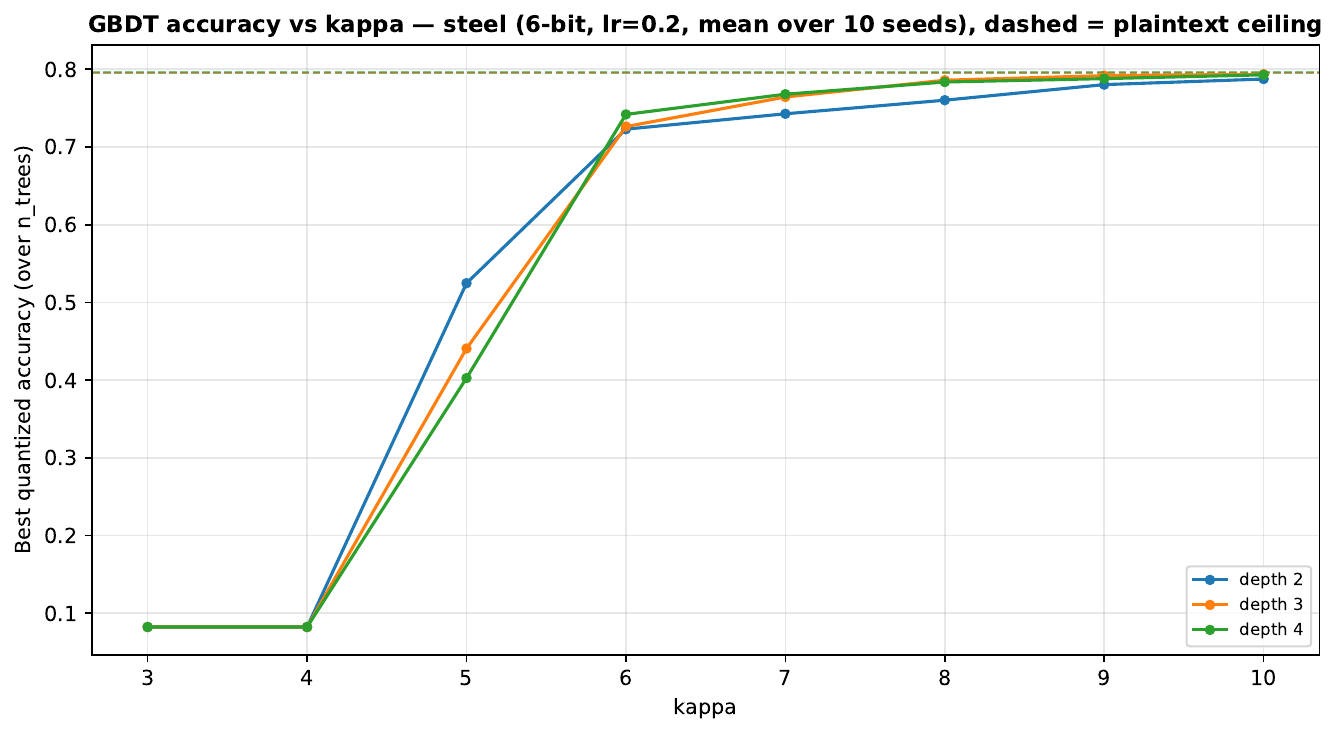}
    \caption{Model accuracy as a function of the rescaling factor \(\kappa\) for gradient boosting forests of depths \(2\)--\(4\). For \(\kappa \le 6\), accuracy degrades substantially; for \(\kappa \ge 9\), the quantized accuracy of depths \(3\) and \(4\) converges to the plaintext baseline. A learning rate of \(0.2\) is chosen to balance model accuracy and computational efficiency.}
    \label{fig:gbdt_kappa}
\end{figure}

\begin{figure}[t]
    \centering
    \includegraphics[width=\columnwidth]{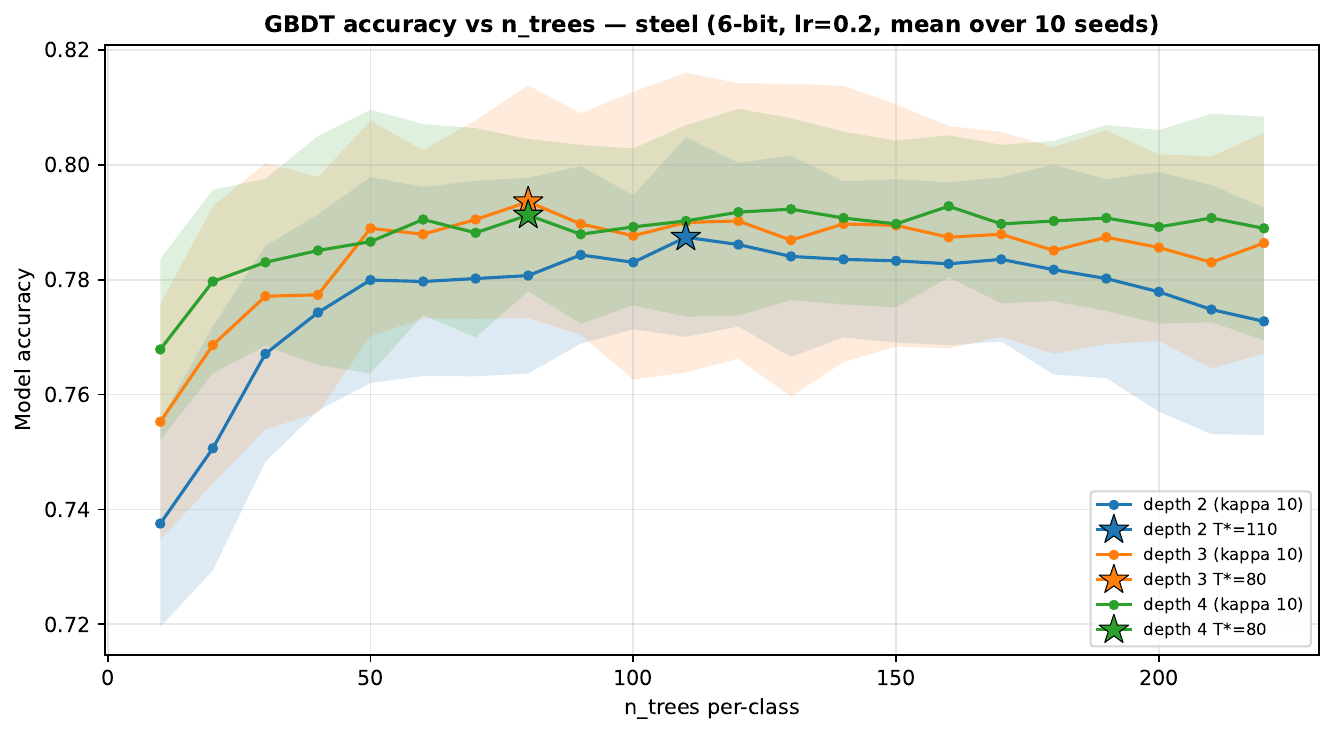}
    \caption{Model accuracy as a function of the per-class tree count for gradient boosting forests of depths \(2\)--\(4\). Each curve is evaluated at the largest rescaling factor \(\kappa\) that achieves accuracy within \(0.01\) of the plaintext ceiling. Depth \(3\) weakly dominates depth \(4\) on cost with no statistical accuracy sacrifice.}
    \label{fig:gbdt_acc}
\end{figure}

\subsection{Extension to Gradient Boosting Ensembles}\label{subsec:ensemble}
The extension to GBDT is straightforward for both our method and the baseline, as the ensemble output is simply the sum of the individual trees' leaf weights~\cite{KMYO+24,DABY25}.
Each tree is evaluated independently, and its packed leaf selection returns a single TRLWE ciphertext carrying the signed-integer weight of the active leaf at coefficient zero.
Per-tree sample extraction and key switching can be deferred and applied after grouping the tree outputs to reduce both computational and noise overhead.
The ensemble prediction is then formed by summing these \(T\) ciphertexts via the additive homomorphism of TRLWE, followed by a conversion to TLWE. Consequently, the per-tree leaf-selection cost is reduced exactly as in the single-tree case from \(|\mathcal{L}|\) to \(|\mathcal{P}|\) PBSs, so the dominant cost of the entire ensemble falls by the same factor.

Even the largest single-tree plaintext modulus (\(p=32\), reached at \(\eta=4\)) is insufficient to hold the aggregated result of an \(80\) trees per class forest for the \(7\)-class Steel dataset.
To overcome this limitation, we encode the score on a finer torus grid of step \(2^{-\kappa} \cdot q\), decoupled from the input plaintext modulus.
Because the leaf weights are quantized according to \(\kappa\), the model accuracy is directly governed by this parameter, as shown in Figure~\ref{fig:gbdt_kappa}.
This design choice demands more careful parameter selection, which we elaborate in Section~\ref{subsec:discuss-param-gbdt}.

\begin{table}[t]
    \centering
    \small
    \caption{GBDT inference time and speedup at depth 3 with 80 trees per class on the steel dataset.}
    \label{tab:gbdt}
    \begin{tabular}{lcc}
        \toprule
        Method & Time (s) & Speedup \\
        \midrule
        OURS   & 13.5     & \(\mathbf{2\times}\) \\
        TE0    & 27.5     & \(1\times\) \\
        \bottomrule
    \end{tabular}
\end{table}

Our experiment on the 7-class Steel dataset (Table~\ref{tab:gbdt}) confirms a \(\SpeedupGbdt\times\) speedup for a depth \(3\) gradient boosting forest comprising \(80\) trees per class\footnote{Optimal model over depth \(2\)--\(4\) and \(10\)--\(220\) trees per class (Figure~\ref{fig:gbdt_acc}). We omit depth \(1\) models here since it is too shallow for the Steel task.}, \footnote{The speedup is insensitive to the thread budget: measured at \(t = 1, 2, 4, 8, 16, 32\), it remains within \(1.94\times\)--\(2.07\times\), closely matching the blind-rotation ratio of \(|\mathcal{L}|/|\mathcal{P}| = 1.97\).}, validating the reduction from \(|\mathcal{L}|\) to \(|\mathcal{P}|\) PBSs. The encrypted output agrees with the quantized-plaintext ensemble on every test instance, consistent with the \(\leq 2^{-64}\) bound.
The ensemble raises model accuracy from the single tree's \(0.61\) to \(0.79\), confirming the effectiveness of gradient boosting.

We defer the per-class argmax operation to the client, as server-side argmax constitutes a separate line of work.
Accordingly, for the Steel dataset, the server returns \(7\) TLWE ciphertexts (28 KB in total, still an order of magnitude below the query's 864 KB), one per class, to the client.

%% file: discussion.tex
\subsection{Parameter Choice for Gradient Boosting Ensemble}\label{subsec:discuss-param-gbdt}
The failure analysis extends to ensembles in a straightforward manner.
Condition~\eqref{eq:cond-in} is enforced by a union bound over all \(T \cdot |\mathcal{P}|\) blind rotations, while the output condition is applied once to the aggregated ciphertext, whose noise variance grows linearly in \(T\), giving
\[
    p_{\mathrm{error}} \leq T |\mathcal{P}| \operatorname{erfc}\!\left(\tfrac{m_{in}}{\sigma_{\mathrm{in}}\sqrt{2}}\right)
    + \operatorname{erfc}\!\left(\tfrac{m_{out}}{\sigma_{\mathrm{out}}\sqrt{2}}\right),
\]
with \(\sigma^2_{\mathrm{out}} = T |\mathcal{P}| V_{\mathrm{br}} + V_{\mathrm{se+ks}}\) and \(m_{\mathrm{out}}=2^{-\kappa}/2\) on the score grid.

Selecting an appropriate rescaling factor \(\kappa\) involves balancing two constraints that pull in opposite directions.
From below, \(\kappa\) must be large enough for the quantized leaf weights to preserve model accuracy: as Figure~\ref{fig:gbdt_kappa} shows, accuracy collapses at \(\kappa = 3,4\) and recovers gradually through \(\kappa = 10\), with depth \(2\) ensembles requiring \(\kappa = 12\) to match the plaintext ceiling, which lies outside the noise-feasible window.
From above, \(\kappa\) is bounded by the available torus budget.
Summing \(T\) trees accumulates the noise.
At the target failure probability (here \(2^{-64}\)), it consumes a fixed number of least-significant bits.
The torus width must therefore satisfy \( \text{torus bits} \ge \kappa - 1 + \text{noise LSB} \).
Under this condition, the parameter choice \(q_{\mathrm{rlwe}} = 2^{32}\) proves insufficient for the Steel model.
Consequently, we adopt \(q_{\mathrm{rlwe}} = 2^{56}\) and \(\kappa = 10\), as this configuration achieves the highest model accuracy among the candidates examined while simultaneously meeting the correctness requirement \(p_{\mathrm{error}} \leq 2^{-64}\).

\subsection{Orthogonality to the Comparison Stage}\label{subsec:discuss-cmp}
The saving we obtain is confined primarily to the PBS count of the leaf-selection stage and is independent of how the per-node branch indicator \(\varepsilon\) is produced.
The comparator is a black box that returns \(\varepsilon\), so any homomorphic comparison operator may be substituted without altering the leaf-selection primitive or its correctness.
We employ the polynomial evaluation version of XCMP because homomorphic comparison is not the central focus of this work.
More recent comparators, such as the XXCMP, RCC~\cite{MNLK23,CKNP24} and the RDCMP~\cite{LLG+24}, offer improved precision or throughput and serve as drop-in replacements that leave our leaf-selection contribution unaffected.
Consequently, our reported results constitute a lower bound on what the encoding can deliver in a fully optimized pipeline.

\subsection{Interaction with Tree Truncation}\label{subsec:discuss-trunc}
Tree truncation is a leaf-level pruning that acts as a plaintext preprocessing step.
It is independent of the query and applies to both schemes.
For example, the truncation of~\cite{CKNP24} relabels the most frequent class as zero and removes zero-valued leaves from leaf selection.
Those leaves contribute nothing to the aggregated output.
Both schemes already reserve that value.
An off-path leaf-parent extracts zero in our encoding, and a leaf whose bucket is not selected extracts zero in TE0.
For TE0, the relabeling removes every leaf of the most frequent class.
For our encoding, it drops a leaf-parent only when both of its children are of that class.
A leaf-parent with one such child still requires its PBS, with an empty bucket in its LUT.
The truncation transfers to our TFHE realization of TE0 unchanged.
A leaf whose label is zero gives an identically zero test polynomial.
Its PBS is a no-op and can be omitted without altering the aggregate.
Here the saving is confined to the PBS count.
The plaintext modulus must cover the range of the path sums.
The relabeling does not change that range, so it brings no parameter relief.

A node with one leaf child and one internal child costs one PBS in both schemes whenever that leaf survives.
Such nodes therefore cancel in the comparison.
Now take the nodes with two leaf children.
Write \(a\), \(b\) and \(c\) for how many of them keep two, one and no non-zero leaf under truncation.
On these nodes, the baseline pays \(2a+b\) PBSs and our encoding \(a+b\), a difference of exactly \(a\).
Truncation can therefore bring the baseline down to our count, but never below it.
So our encoding is never worse than the per-leaf baseline.
The speedup is at most \(1+a/(a+b)\), with equality on a tree that has no single-leaf node.
As in the untruncated comparison of Section~\ref{subsec:arbitrary}, the ratio stays between one and two.
This bound equals one when no node keeps two non-zero leaf children.
One such tree is a binary classification tree whose sibling leaves always carry distinct labels.
It approaches the full factor of two for multi-class trees and for gradient-boosted ensembles, whose quantized leaf scores are rarely exactly zero.
Truncation also makes the PBS count depend on the label distribution, which is acceptable since the model is not treated as confidential.

\subsection{Extension to j-ary Trees}\label{subsec:discuss-nary}
Our path encoding extends naturally to trees with \(j\)-ary splits, since it consumes a multivalued routing outcome natively.
For a \(j\)-ary node, a single comparison with a \(j\)-interval polynomial node outputs the branch index \(d \in \{0,\dots,j-1\}\), which is then scaled by a plaintext radix weight and accumulated into the path sum and resolved by a single PBS at each leaf-parent.
The LUT is encoded with the true leaf labels at different plaintext slots as in the binary case.
By contrast, state-of-the-art encodings assume the binary tree, assigning the selected node with a value of 0 (zero-sum) or 1 (path-conjugation), lacking the ability to distinguish multi-nary nodes.

Two aspects of the binary construction do not transfer automatically and are left open.
First, our optimality results are limited to binary-valued variables.
The corresponding minimal-weight question for \(j\)-valued variables is, to our knowledge, open.
Second, the cost of this generality is message range.
The path sum occupies \(\eta \lceil \log_2 j \rceil\) bits of plaintext space, and since the radix weights are plaintext multipliers, the comparison noise scales with \(j^{2i}\) at level \(i\).
Both effects tighten the feasibility window of Section~\ref{subsec:discuss-param-gbdt}.
We therefore state the extension as an analytical property of the encoding rather than an evaluated result.

%% file: main.bbl
\begin{thebibliography}{10}
\providecommand{\url}[1]{#1}
\csname url@samestyle\endcsname
\providecommand{\newblock}{\relax}
\providecommand{\bibinfo}[2]{#2}
\providecommand{\BIBentrySTDinterwordspacing}{\spaceskip=0pt\relax}
\providecommand{\BIBentryALTinterwordstretchfactor}{4}
\providecommand{\BIBentryALTinterwordspacing}{\spaceskip=\fontdimen2\font plus
\BIBentryALTinterwordstretchfactor\fontdimen3\font minus
  \fontdimen4\font\relax}
\providecommand{\BIBforeignlanguage}[2]{{%
\expandafter\ifx\csname l@#1\endcsname\relax
\typeout{** WARNING: IEEEtran.bst: No hyphenation pattern has been}%
\typeout{** loaded for the language `#1'. Using the pattern for}%
\typeout{** the default language instead.}%
\else
\language=\csname l@#1\endcsname
\fi
#2}}
\providecommand{\BIBdecl}{\relax}
\BIBdecl
\renewcommand{\BIBentryALTinterwordstretchfactor}{4}

\bibitem{GOV22}
L{\'e}o Grinsztajn, Edouard Oyallon, and Ga{\"e}l Varoquaux, ``Why do
  tree-based models still outperform deep learning on typical tabular data?''
  \emph{Advances in neural information processing systems}, vol.~35, pp.
  507--520, 2022.

\bibitem{BPTG15}
\BIBentryALTinterwordspacing
Raphael Bost, Raluca~Ada Popa, Stephen Tu, and Shafi Goldwasser, ``Machine
  learning classification over encrypted data,'' Cryptology {ePrint} Archive,
  Paper 2014/331, 2014. Available: \url{https://eprint.iacr.org/2014/331}
\BIBentrySTDinterwordspacing

\bibitem{WFNL16}
David~J Wu, Tony Feng, Michael Naehrig, and Kristin Lauter, ``Privately
  evaluating decision trees and random forests,'' \emph{Proceedings on Privacy
  Enhancing Technologies}, 2016.

\bibitem{TMZC17}
Raymond~KH Tai, Jack~PK Ma, Yongjun Zhao, and Sherman~SM Chow,
  ``Privacy-preserving decision trees evaluation via linear functions,'' in
  \emph{European Symposium on Research in Computer Security}.\hskip 1em plus
  0.5em minus 0.4em\relax Springer, 2017, pp. 494--512.

\bibitem{KNLA19}
{\'A}gnes Kiss, Masoud Naderpour, Jian Liu, N~Asokan, and Thomas Schneider,
  ``Sok: Modular and efficient private decision tree evaluation,''
  \emph{Proceedings on Privacy Enhancing Technologies}, 2019.

\bibitem{MTZC21}
Jack~PK Ma, Raymond~KH Tai, Yongjun Zhao, and Sherman~SM Chow, ``Let’s stride
  blindfolded in a forest: Sublinear multi-client decision trees evaluation,''
  in \emph{The Network and Distributed System Security Symposium (NDSS)}, 2021.

\bibitem{ADD22}
Sona Alex, KJ~Dhanaraj, and PP~Deepthi, ``Private and energy-efficient decision
  tree-based disease detection for resource-constrained medical users in mobile
  healthcare network,'' \emph{IEEE Access}, vol.~10, pp. 17\,098--17\,112,
  2022.

\bibitem{YLQ+24}
Shuai Yuan, Hongwei Li, Xinyuan Qian, Meng Hao, Yixiao Zhai, and Guowen Xu,
  ``Efficient and privacy-preserving outsourcing of gradient boosting decision
  tree inference,'' \emph{IEEE Transactions on Services Computing}, vol.~17,
  no.~5, pp. 2334--2348, 2024.

\bibitem{KMYO+24}
\BIBentryALTinterwordspacing
Ronny Ko, Abdelkarim Kati, Robin Geelen, Rasoul~Akhavan Mahdavi, Byoungwoo
  Yoon, Jongho Shin, Zhiqiang Lin, Makoto Onizuka, and Florian Kerschbaum,
  ``Silentwood: Efficient private inference over gradient-boosting decision
  forests,'' 2026. Available: \url{https://arxiv.org/abs/2411.15494}
\BIBentrySTDinterwordspacing

\bibitem{FWO25}
Satoshi Fukui, Lihua Wang, and Seiichi Ozawa, ``Efficient and
  privacy-preserving decision tree inference via homomorphic matrix
  multiplication and leaf node pruning,'' \emph{Applied Sciences}, vol.~15,
  no.~10, p. 5560, 2025.

\bibitem{XZZ+26}
\BIBentryALTinterwordspacing
Wei Xu, Hui Zhu, Yandong Zheng, Song Bian, Ning Sun, Yuan Hao, Dengguo Feng,
  and Hui Li, ``Kangaroo: A private and amortized inference framework over wan
  for large-scale decision tree evaluation,'' in \emph{Proceedings 2026 Network
  and Distributed System Security Symposium}, ser. NDSS 2026.\hskip 1em plus
  0.5em minus 0.4em\relax Internet Society, 2026. Available:
  \url{http://dx.doi.org/10.14722/ndss.2026.230892}
\BIBentrySTDinterwordspacing

\bibitem{LZS18}
Wen-jie Lu, Jun-Jie Zhou, and Jun Sakuma, ``Non-interactive and output
  expressive private comparison from homomorphic encryption,'' in
  \emph{Proceedings of the 2018 on Asia Conference on Computer and
  Communications Security}, 2018, pp. 67--74.

\bibitem{XWCF19}
Xiaodong Xiao, Ting Wu, Yuanfang Chen, and Xingyue Fan, ``Privacy-preserved
  approximate classification based on homomorphic encryption,''
  \emph{Mathematical and Computational Applications}, vol.~24, no.~4, p.~92,
  2019.

\bibitem{TBK20}
Anselme Tueno, Yordan Boev, and Florian Kerschbaum, ``Non-interactive private
  decision tree evaluation,'' in \emph{IFIP Annual Conference on Data and
  Applications Security and Privacy}.\hskip 1em plus 0.5em minus 0.4em\relax
  Springer, 2020, pp. 174--194.

\bibitem{SNR+21}
\BIBentryALTinterwordspacing
Kanthi Sarpatwar, Karthik Nandakumar, Nalini Ratha, James Rayfield, Karthikeyan
  Shanmugam, Sharath Pankanti, and Roman Vaculin, ``Efficient encrypted
  inference on ensembles of decision trees,'' 2021. Available:
  \url{https://arxiv.org/abs/2103.03411}
\BIBentrySTDinterwordspacing

\bibitem{ADGK22}
Sofiane Azogagh, Victor Delfour, S{\'e}bastien Gambs, and Marc-Olivier
  Killijian, ``Probonite: Private one-branch-only non-interactive decision tree
  evaluation,'' in \emph{Proceedings of the 10th Workshop on Encrypted
  Computing \& Applied Homomorphic Cryptography}, 2022, pp. 23--33.

\bibitem{CDPP22}
Kelong Cong, Debajyoti Das, Jeongeun Park, and Hilder~VL Pereira, ``Sortinghat:
  Efficient private decision tree evaluation via homomorphic encryption and
  transciphering,'' in \emph{Proceedings of the 2022 ACM SIGSAC Conference on
  Computer and Communications Security}, 2022, pp. 563--577.

\bibitem{PTVA22}
Jestine Paul, Benjamin Hong~Meng Tan, Bharadwaj Veeravalli, and Khin Mi~Mi
  Aung, ``Non-interactive decision trees and applications with multi-bit
  tfhe,'' \emph{Algorithms}, vol.~15, no.~9, p. 333, 2022.

\bibitem{MNLK23}
Rasoul Akhavan~Mahdavi, Haoyan Ni, Dimitry Linkov, and Florian Kerschbaum,
  ``Level up: Private non-interactive decision tree evaluation using levelled
  homomorphic encryption,'' in \emph{Proceedings of the 2023 ACM SIGSAC
  Conference on Computer and Communications Security}, 2023, pp. 2945--2958.

\bibitem{FSB+23}
Jordan Frery, Andrei Stoian, Roman Bredehoft, Luis Montero, Celia Kherfallah,
  Benoit Chevallier-Mames, and Arthur Meyre, ``Privacy-preserving tree-based
  inference with tfhe,'' in \emph{International Conference on Mobile, Secure,
  and Programmable Networking}.\hskip 1em plus 0.5em minus 0.4em\relax
  Springer, 2023, pp. 139--156.

\bibitem{CKNP24}
Kelong Cong, Jiayi Kang, Georgio Nicolas, and Jeongeun Park, ``Faster private
  decision tree evaluation for batched input from homomorphic encryption,'' in
  \emph{International Conference on Security and Cryptography for
  Networks}.\hskip 1em plus 0.5em minus 0.4em\relax Springer, 2024, pp. 3--23.

\bibitem{LLG+24}
Huiqiang Liang, Haining Lu, Yifeng Guo, Geng Wang, Haining Yu, Hongli Zhang,
  Baoyu An, Jinyu Li, and Li~Su, ``Bpdte: Batch private decision tree
  evaluation via amortized efficient private comparison,'' \emph{Cryptology
  ePrint Archive, Paper 2024/619}, 2024.

\bibitem{DABY26}
El~Hadji~Mamadou Dia, Walid Arabi, Anis Bkakria, and Reda Yaich, ``Strawhat:
  Private non-interactive gradient boosting decision tree evaluation based on
  homomorphic encryption,'' in \emph{International Symposium on Foundations and
  Practice of Security}.\hskip 1em plus 0.5em minus 0.4em\relax Springer, 2025,
  pp. 40--59.

\bibitem{XTWA23}
Kexin Xu, Benjamin Hong~Meng Tan, Li-Ping Wang, Khin Mi~Mi Aung, and Huaxiong
  Wang, ``Privacy-preserving outsourcing decision tree evaluation from
  homomorphic encryption,'' \emph{Journal of Information Security and
  Applications}, vol.~77, p. 103582, 2023.

\bibitem{PCL26}
\BIBentryALTinterwordspacing
Dongjin Park, Gyeongwon Cha, and Joon-Woo Lee, ``{LCPDTE}: Low-complexity
  private decision tree evaluation over homomorphic encryption,'' Cryptology
  {ePrint} Archive, Paper 2026/1263, 2026. Available:
  \url{https://eprint.iacr.org/2026/1263}
\BIBentrySTDinterwordspacing

\bibitem{DABY25}
DIA El~Hadji~Mamadou, Walid ARABI, Anis BKAKRIA, and Reda YAICH, ``A
  comprehensive survey of privacy-preserving decision trees based on
  homomorphic encryption,'' \emph{Cryptology ePrint Archive, Paper 2025/1310},
  2025.

\bibitem{SCL+24}
Hojune Shin, Jina Choi, Dain Lee, Kyoungok Kim, and Younho Lee, ``Fully
  homomorphic training and inference on binary decision tree and random
  forest,'' in \emph{European Symposium on Research in Computer
  Security}.\hskip 1em plus 0.5em minus 0.4em\relax Springer, 2024, pp.
  217--237.

\bibitem{CGGI20}
\BIBentryALTinterwordspacing
Ilaria Chillotti, Nicolas Gama, Mariya Georgieva, and Malika Izabach{\`e}ne,
  ``Tfhe: Fast fully homomorphic encryption over the torus,'' \emph{J.
  Cryptol.}, vol.~33, no.~1, pp. 34--91, Jan. 2020. Available:
  \url{https://doi.org/10.1007/s00145-019-09319-x}
\BIBentrySTDinterwordspacing

\bibitem{TZJR+16}
Florian Tram\`{e}r, Fan Zhang, Ari Juels, Michael~K. Reiter, and Thomas
  Ristenpart, ``Stealing machine learning models via prediction apis,'' in
  \emph{Proceedings of the 25th USENIX Conference on Security Symposium}, ser.
  SEC'16.\hskip 1em plus 0.5em minus 0.4em\relax USA: USENIX Association, 2016,
  pp. 601--618.

\bibitem{BB19}
Brad Boehmke and Brandon Greenwell, \emph{Hands-On Machine Learning with
  R}.\hskip 1em plus 0.5em minus 0.4em\relax Chapman and Hall/CRC, 11 2019.

\bibitem{APS15}
\BIBentryALTinterwordspacing
Martin~R. Albrecht, Rachel Player, and Sam Scott, ``On the concrete hardness of
  learning with errors,'' \emph{Journal of Mathematical Cryptology}, vol.~9,
  no.~3, pp. 169--203, 2015. Available:
  \url{https://doi.org/10.1515/jmc-2015-0016}
\BIBentrySTDinterwordspacing

\bibitem{BCCC+25}
Jean-Philippe Bossuat, Rosario Cammarota, Ilaria Chillotti, Benjamin~R. Curtis,
  Wei Dai, Huijing Gong, Erin Hales, Duhyeong Kim, Bryan Kumara, Changmin Lee,
  Xianhui Lu, Carsten Maple, Alberto Pedrouzo-Ulloa, Rachel Player, Yuriy
  Polyakov, Luis Antonio~Ruiz Lopez, Yongsoo Song, and Donggeon Yhee,
  ``Security guidelines for implementing homomorphic encryption,'' \emph{{IACR}
  Communications in Cryptology}, vol.~1, no.~4, 2025.

\bibitem{BJSW25}
Olivier Bernard, Marc Joye, Nigel~P. Smart, and Michael Walter, ``Drifting
  towards better error probabilities in fully homomorphic encryption
  schemes,'' in \emph{Advances in Cryptology -- EUROCRYPT 2025}, Serge Fehr and
  Pierre-Alain Fouque, Eds.\hskip 1em plus 0.5em minus 0.4em\relax Cham:
  Springer Nature Switzerland, 2025, pp. 181--211.

\end{thebibliography}
